%% file: main.tex
\documentclass[10pt,journal,compsoc]{IEEEtran}

\ifCLASSOPTIONcompsoc
  \usepackage[nocompress]{cite}
\else
  \usepackage{cite}
\fi

\usepackage{booktabs} % For formal tables
\usepackage[caption=false]{subfig}
\usepackage{graphicx}
\usepackage{epstopdf}
\usepackage{xcolor}
\usepackage{url}
\usepackage{cite}
\usepackage{hyperref}

\begin{document}
\title{COUNTDOWN: a Run-time Library for Performance-Neutral Energy Saving\\in MPI Applications}
%\title{{\color{red}COUNTDOWN: a Run-time Library for Energy Saving in MPI-based Scientific Applications}}

\author{Daniele~Cesarini,
        Andrea~Bartolini,~\IEEEmembership{Member,~IEEE,}
        Pietro~Bonf\`a,\\
        Carlo~Cavazzoni
        and~Luca~Benini,~\IEEEmembership{Fellow,~IEEE}
        
%\thanks{Manuscript received January 30, 2014; This work was supported by NSF Variability Expeditions (1029783), ERC-AdG MultiTherman (291125), and FP7 Virtical (288574).}% <-this % stops a space

\IEEEcompsocitemizethanks{\IEEEcompsocthanksitem D. Cesarini and A. Bartolini are with the Department of Electrical, Electronic and Information Engineering "Guglielo Marconi", University of Bologna, 40136 Bologna, Italy (e-mail: daniele.cesarini@unibo.it;a.bartolini@unibo.it).
\IEEEcompsocthanksitem P. Bonf\`a and C. Cavazzoni are with the Department of SuperComputing Applications and Innovation, CINECA, 40033 Casalecchio di Reno (BO), Italy (e-mail: p.bonfa@cineca.it;c.cavazzoni@cineca.it).
\IEEEcompsocthanksitem L. Benini is with the Department of Information Technology and Electrical Engineering, Swiss Federal Institute of Technology in Zurich, 8092 Zurich, Switzerland and
the Department of Electrical, Electronic and Information Engineering "Guglielo Marconi", University of Bologna, 40136 Bologna, Italy (e-mail: lbenini@iis.ee.ethz.ch).}
}

%{Shell \MakeLowercase{\textit{et al.}}: Bare Demo of IEEEtran.cls for Computer Society Journals}

% apply IEEE BibTeX customizations
%\bstctlcite{IEEEexample:BSTcontrol}

\maketitle

\IEEEdisplaynontitleabstractindextext
\IEEEpeerreviewmaketitle

%\IEEEtitleabstractindextext{
\begin{abstract}
%Power consumption is a looming treat in today's computing progress. In scientific computing, a significant amount of power is spent in the communication and synchronization-related idle times.
%Power consumption is today the key limiting factor in high-performance computing (HPC). 
Power and energy consumption is becoming key challenges to deploy the first exascale supercomputer successfully.
Large-scale HPC applications waste a significant amount of power in communication and synchronization-related idle times.
However, due to the time scale at which communication happens, transitioning in low power states during communication's idle times may introduce unacceptable overhead in applications' execution time.

In this paper, we present COUNTDOWN, a runtime library, supported by a methodology and analysis tool for identifying and automatically reducing the power consumption of the computing elements during communication and synchronization.
COUNTDOWN saves energy without imposing significant time-to-completion increase by lowering CPUs power consumption only during idle times for which power state transition overhead are negligible. 
%primitives filtering out phases which would detriment the time to solution of the application. 
This is done transparently to the user, without requiring labor-intensive and error-prone application code modifications, nor requiring recompilation of the application. We test our methodology in a production Tier-0 system. For the NAS benchmarks, COUNTDOWN saves between 6\% and 50\% energy, with a time-to-solution penalty lower than 5\%. In a complete production --- Quantum ESPRESSO --- for a 3.5K cores run, COUNTDOWN saves 22.36\% energy, with a performance penalty below 3\%.
Energy saving increases to 37\% with a performance penalty of 6.38\%, if the application is executed without communication tuning.

%Power ad energy consumption are becoming key challenges to face in order to reach the first exascale supercomputer.
%State-of-the-art research in power and energy saving for high performance computing relies on DVFS techniques to reduce cores' frequency to adjust the workload imbalance of the application predicting the evolution of the application.
%However, mispredictions can happen in irregular applications which cause performance penalty, moreover applications must manifest unbalance workload in order to prioritize processes.
%In this paper, we present EPEO, a runtime system for fine-grain profiling and energy reduction for HPC applications which leverage on communication slacks.
%Our methodology is purely reactive and it is based on timer strategies to target communication primitives with high energy consumption leaving unmodified system's performance in application workload.
%We tested our methodology in a production HPC system using two widely HPC code, Quantum ESPRESSO (QE) and Gromacs (GR), which are real-life workload scientific applications.
%We use production datasets for our benchmarks which can scale up to 4K cores of the system using tens of compute nodes.
%We proved that our proposed methodology can reach up to 24\% of energy saving with a performance gain of 8\%.
\end{abstract}

%}

\IEEEdisplaynontitleabstractindextext
\IEEEpeerreviewmaketitle

\input{introduction}

\input{related}

\input{background}

\input{framework}

\input{experimental}

\input{conclusion}

\ifCLASSOPTIONcompsoc
  \section*{Acknowledgments}
\else
  \section*{Acknowledgment}
\fi
Work supported by the EU FETHPC project ANTAREX (g.a. 671623), EU project ExaNoDe (g.a. 671578), and CINECA research grant on Energy-Efficient HPC systems.

%\bibliographystyle{IEEEtran}
\input{main.bbl}

%\bibliography{main}

\vskip -2.5\baselineskip plus -10fil

\begin{IEEEbiography}[{\includegraphics[width=1in, height=1.25in, clip, keepaspectratio]{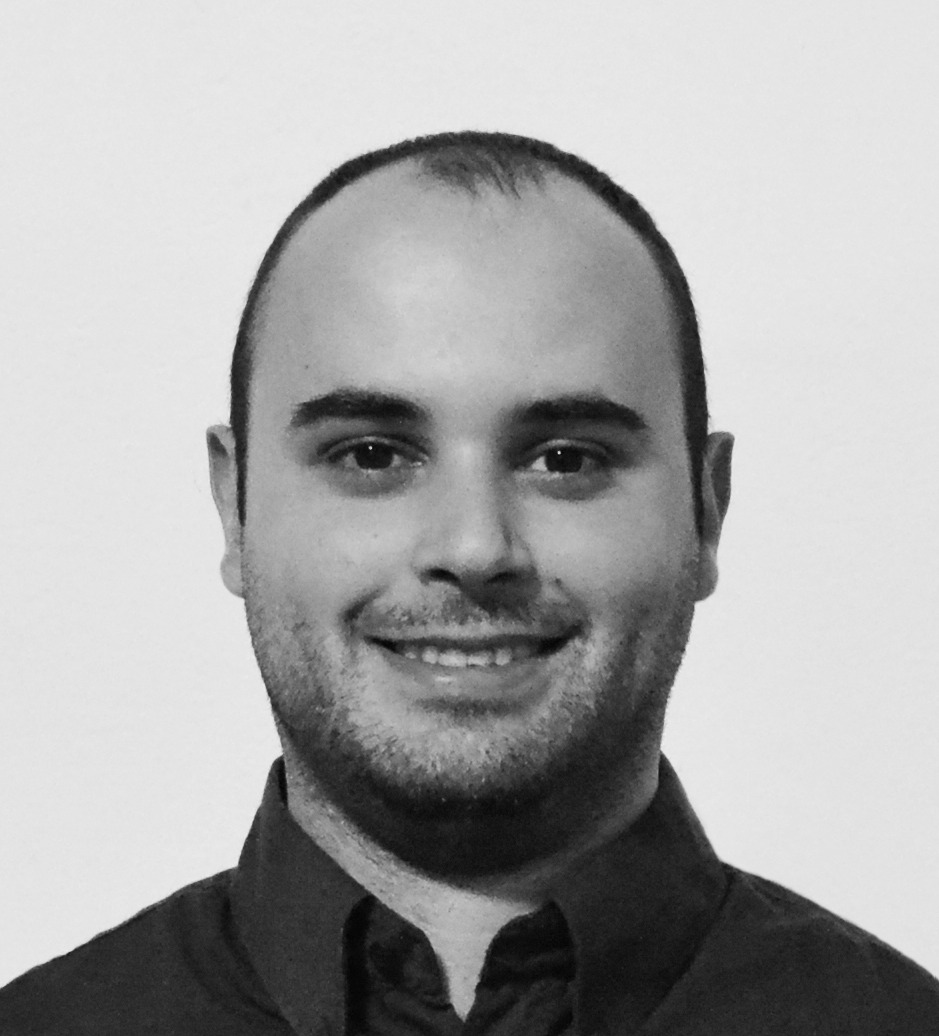}}]{Daniele Cesarini}
received a Ph.D. degree in Electrical Engineering from the University of Bologna, Italy, in 2019, where he is currently a Post-Doctoral researcher in the Department of Electrical, Electronic and Information Engineering (DEI). His research interests concern the development of SW-HW codesign strategies as well as algorithms for parallel programming support for energy-efficient HPC systems.
\end{IEEEbiography}

\vskip -3.5\baselineskip plus -10fil

\begin{IEEEbiography}[{\includegraphics[width=1in, height=1.25in, clip, keepaspectratio]{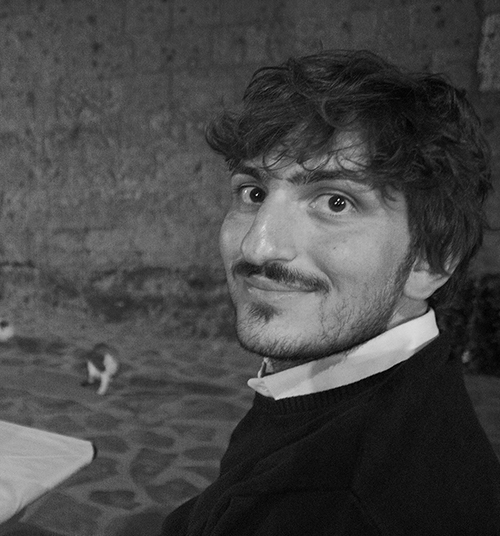}}]{Andrea Bartolini} 
received a Ph.D. degree in Electrical Engineering from the University of Bologna, Italy, in 2011. He is currently Assistant Professor in the Department of Electrical, Electronic and Information Engineering (DEI) at the University of Bologna. Before, he was Post-Doctoral researcher in the Integrated Systems Laboratory at ETH Zurich. Since 2007 Dr. Bartolini has published more than 80 papers in peer-reviewed international journals and conferences with focus on dynamic resource management for embedded and HPC systems. %In latest two years he has led the design team responsible for the energy-efficiency and monitoring extensions in the first Top500 OpenPOWER supercomputer.
\end{IEEEbiography}

\vskip -2.6\baselineskip plus -10fil

\begin{IEEEbiography}[{\includegraphics[width=1in, height=1.25in, clip, keepaspectratio]{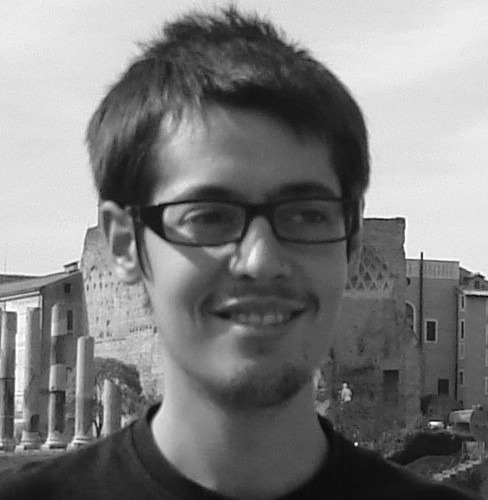}}]{Pietro Bonf\`a}
Dr. Pietro Bonf\`a received his B.Sc. degree in Physical Engineering (2008) in Politecnico di Milano, the M.Sc. degree in Physics (2011) from University of Pavia and the Ph.D. in Physics (2015) from University of
Parma. He has a background in solid state physics and he is now actively involved in the development of computational chemistry codes as part of his activities in CINECA.
\end{IEEEbiography}

\vskip -3.25\baselineskip plus -10fil

\begin{IEEEbiography}[{\includegraphics[width=1in, height=1.25in, clip, keepaspectratio]{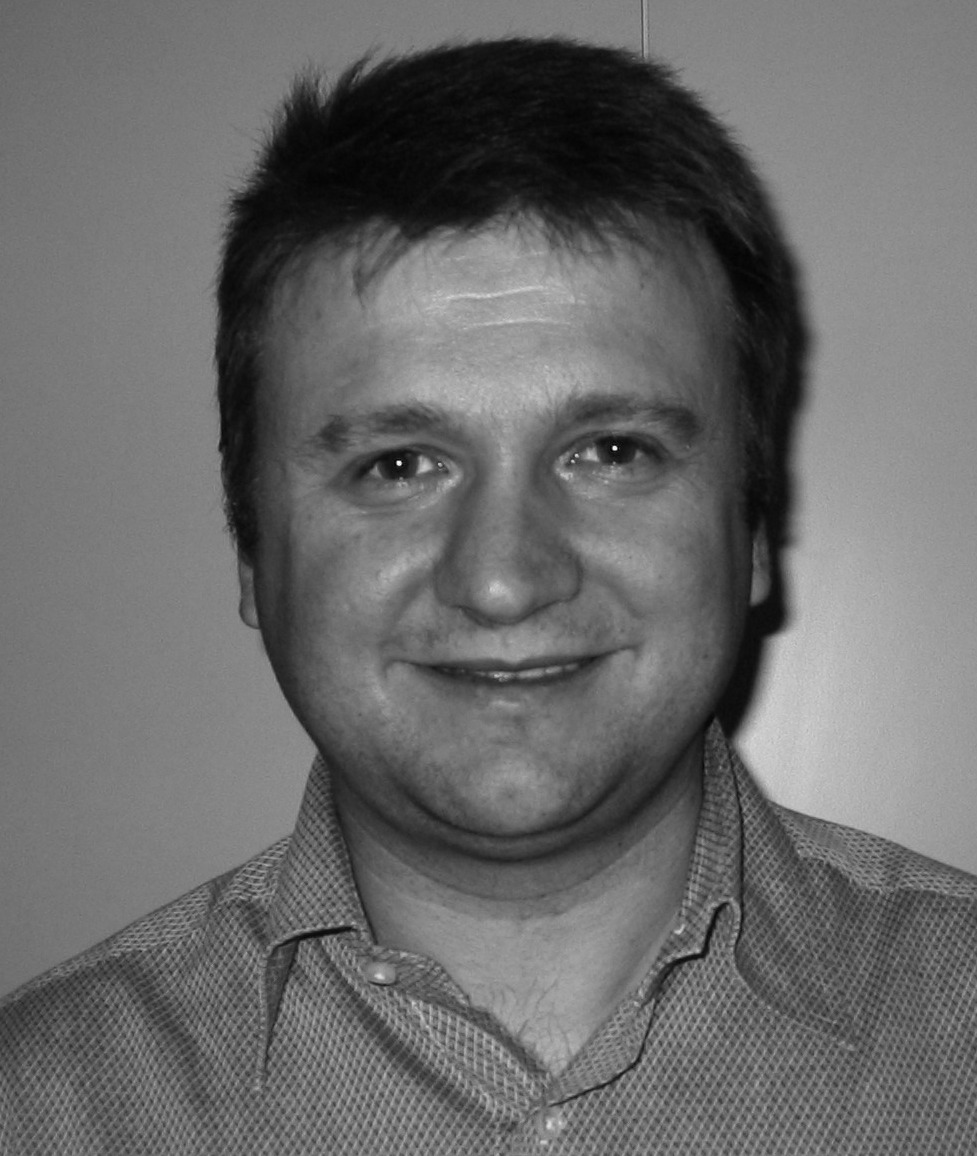}}]{Carlo Cavazzoni}
Carlo graduated cum laude in Physics from the University of Modena and earned his PhD Material Science at the International School for Advanced Studies of Trieste in 1998. 
%He has studied various problems concerning the implementation and the efficiency of parallel numerical algorithms being used in physical computer simulations and 
He has authored or co-authored several papers published in prestigious international review including Science, Physical Review Letters, Nature Materials. Currently in the HPC Business Unit of CINECA, he is responsible for the R\&D, HPC infrastructure evolution and collaborations with scientific communities. 
%He collaborates with different user communities to enable applications on massively parallel systems and innovative architecture solutions. 
%He is responsible for the parallel design of QUANTUM ESPRESSO suite of codes, and for co-design activities in the MaX centre of excellence.
\end{IEEEbiography}

\vskip -2.75\baselineskip plus -10fil

\begin{IEEEbiography}[{\includegraphics[width=1in, height=1.25in, clip, keepaspectratio]{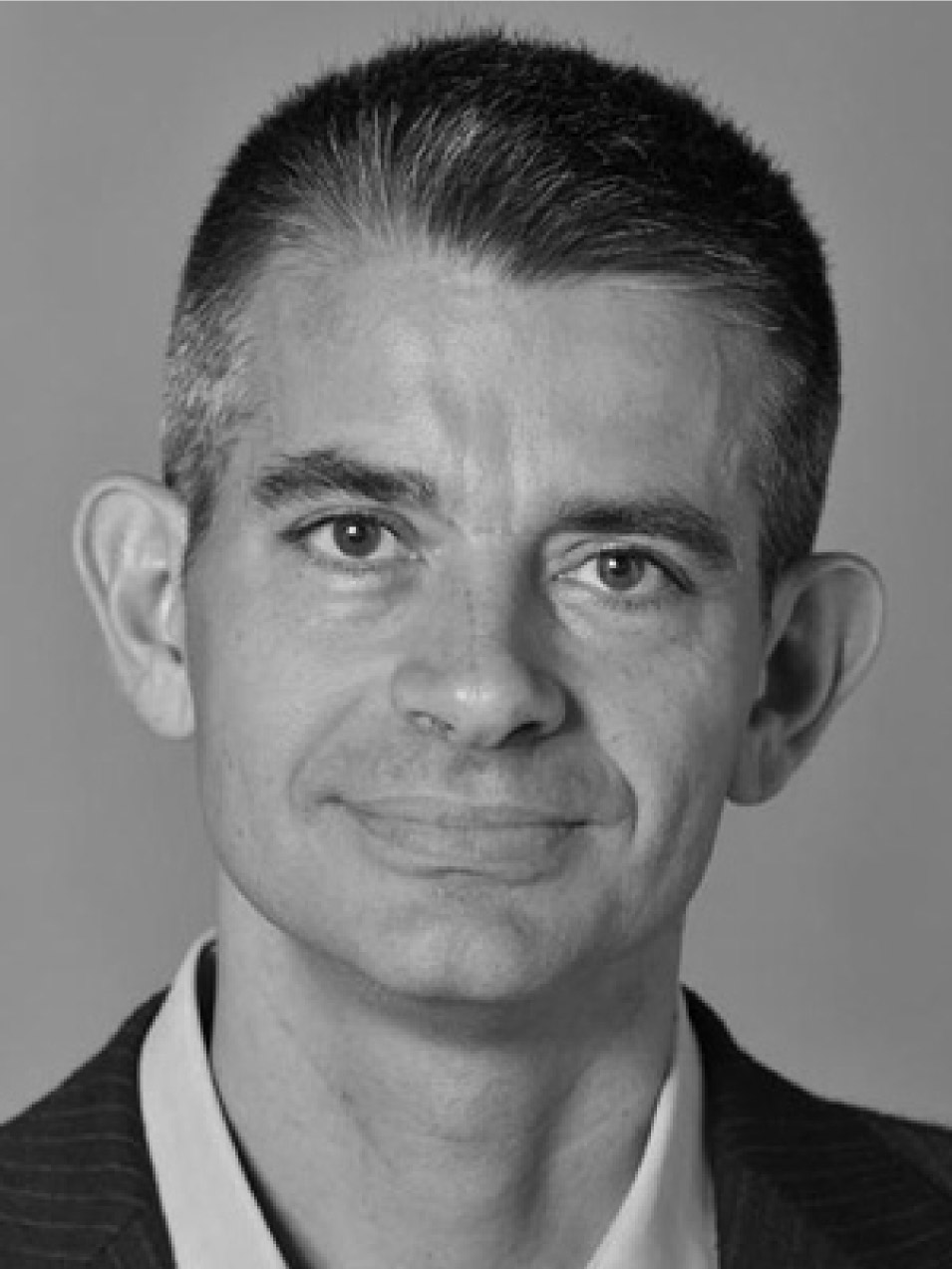}}]{Luca Benini} 
is professor of Digital Circuits and Systems at ETH Zurich, Switzerland, and is also professor at University of Bologna, Italy. His research interests are in system design of energy-efficient multicore SoC, smart sensors and sensor networks. He has published more than 800 papers in peer reviewed international journals and conferences, four books and several book chapters. He is a fellow of the ACM and Member of the Academia Europea. He is the recipient of the IEEE CAS Mac Van Valkenburg Award 2016.

\end{IEEEbiography}

\end{document}

%% file: introduction.tex
\section{Introduction}
\label{sec:introduction}
% Energy and power problem
In today's supercomputers, the total power consumption of computing devices limits practically achievable performance. This is a direct consequence of the end of Dennard's scaling, which in the last decade has caused a progressive increase of the power density required to operate each new processor generation at its maximum performance. Higher power density implies more heat to be dissipated and increases cooling costs. These altogether worsen the total costs of ownership (TCO) and operational costs: limiting de facto the budget for the supercomputer computational capacity.

Low power design strategies enable computing resources to trade-off their performance for power consumption by mean of low power modes of operation. These states obtained by Dynamic and Voltage Frequency Scaling (DVFS) (also known as performance states or P-states \cite{ACPI}), clock gating or throttling states (T-states), and idle states which switch off unused resources (C-states \cite{ACPI}). Power states transitions are controlled by hardware policies, operating system (OS) policies, and with an increasing emphasis in recent years, at user-space by the final users\cite{fraternali_islped04,lrz_lowfreq,losalomos_sc05,GEOPM} and at execution time \cite{adagio_dynamic,Schulz_IPDPS10}.

While OS policies try to maximize the usage of the computing resources --- increasing the processor's speed (P-state) proportionally to the processor's utilization, with a specific focus on server and interactive workload --- two main families of power control policies are emerging in scientific computing. 
The first is based on the assumption that the performance penalty can be tolerated to reduce the overall energy consumption \cite{fraternali_islped04,fraternali_tpds17,lrz_lowfreq,losalomos_sc05}. The second is based on the assumption that it is possible to slow down a processor only when it does not execute critical tasks: to save energy without penalizing application performance \cite{adagio_dynamic,Schulz_IPDPS10,freeh2008just,GEOPM}. Both approaches are based on the concept of application slack/bottleneck (memory, IO, and communication) that can be opportunistically exploited to reduce power and save energy. However, there are drawbacks which limit the usage of these concepts in a production environment. The first approach causes overheads in the application time-to-solution (TTS) limiting the supercomputer throughput and capacity. The second approach depends on the capability of predicting the critical tasks in advance with severe performance loss in case of mispredictions.
% Communication time per salvare energia 

A typical HPC application is composed of several processes running on a cluster of nodes which exchange messages through a high-bandwidth, low-latency network.
%These processes can access the network sub-system through a software interface that abstracts the network level.
%The Message-Passing Interface (MPI) is a simple but high-performance standard interface for communication that allows these application's processes to exchange explicit messages.
These processes can access the network sub-system through a software interface that abstracts the network level.
The Message-Passing Interface (MPI) is a software interface for communication that allows processes to exchange explicit messages abstracting the network level. % through application instances in distributed environments.
%Different supercomputer vendors implement their own MPI library to optimize the communication in network sub-systems.
Usually, when the scale of the application increases, the time spent by the application in the MPI library becomes not negligible and impacts the overall power consumption.
%%% Following paragraph to be written better %%%
By default, when MPI processes are waiting in a synchronization primitive, the MPI libraries use a busy-waiting mechanism. However, during MPI primitives the workload is primarily composed of wait times and IO/memory accesses for which running an application in a low power mode may result in lower CPU power consumption with limited or even no impact on the execution time.

MPI libraries implement idle-waiting mechanisms, but these are not used in practice to avoid performance penalties caused by the transition times into and out of low-power states \cite{hackenberg2015energy}. %Indeed, while these transition time could be tolerable in unbalanced workload with long synchronization primitives they are not for a balanced workload with short MPI synchronization primitives. 
As a matter of fact, there is no known low-overhead and reliable mechanism for reducing energy consumption selectively during MPI communication slack.
%But running an application at a lower frequency in MPI primitive can result in lower CPU power consumption due the natural I\/O and memory boundness of MPI primitives.
%Unfortunately, MPI libraries that implement the possibility to use idle-waiting mechanism do not rely on this functionality to avoid performance penalty.
%Changing the voltage/frequency of a system require privilege permission, instead, HPC applications run at standard user permission, this does not allow MPI libraries to access on DVFS system knobs.
%Countdown è l’utilizzo di un timeout per ridurre gli overhead del trasizione di stati di low power, ma la sua applicazione nel contesto HPC e del saving di energia nell’unbalance. 

%Esiste margine per salvare l’energia senza compromettere le performance in HPC focalizzandosi sull'unbalance
%La capacità di salvare questa energia tramite l'uso di stati di low power è limitata dalle latenze con cui si cambiano gli stati di low power
%In questo lavoro presentiamo countdown, che è al contempo una metodologia ma anche un tool per applicarla.
%Countdown è il fatto di avere una metodologia con timer
% contributions
In this paper, we present COUNTDOWN\footnote{Github Repository: \url{https://github.com/EEESlab/countdown}}, a run-time library, analysis tool, and methodology to save energy in MPI-based applications by leveraging the communication slack. The main contribution of this manuscript are:
%\begin{itemize}

i) An analysis of the effects and implications of fine-grain power management in today's supercomputing systems targeting energy saving in the MPI library. Our study shows that in today's HPC processors there are significant latencies in the HW to serve low power states transitions. We show that this delay is at the source of inefficiencies (overheads and saving losses) in the application for fine-grain power management in the MPI library.

ii) Through the first set of benchmarks running on a single HPC node we show that: (a) there is a potential saving of energy with negligible overheads in the MPI communication slack of today's HPC applications; (b) these savings are jeopardized by the time that HW takes to perform power state transitions; (c) when combined with low-power states, Turbo logic can help improving execution time.

iii) The COUNTDOWN library, which consists of a runtime able to automatically track at fine granularity MPI and application phases to inject power management calls. COUNTDOWN can identify MPI calls with energy-saving potential for which it is worthwhile to enter a low power state, leaving low-wait-time MPI calls unmodified to prevent overheads caused by low power state transitions. We show that COUNTDOWN's principles can be used to inject DVFS calls as well as to configure the MPI runtime correctly and take advantage of MPI idle-waiting mechanisms. COUNTDOWN works at execution time without requiring any off-line knowledge of the application, and it is completely: it does not require any modification of the source code and compilation toolchain. COUNTDOWN can be dynamically linked with the application at loading time: it can intercept dynamic linking to the MPI library instrumenting all the application calls to MPI functions before the execution workflow jumps to the library. The runtime also provides a static version of the library which can be connected with the application at linking time. COUNTDOWN supports C/C++ and Fortran HPC applications and most of the open-source and commercial MPI libraries.

iv) We evaluate COUNTDOWN with a wide set of benchmarks and low power state mechanisms. 
%We show that the COUNTDOWN approach reduces the overheads of fine-grain power management using T-states of 1.70\%, P-states of the 0.02\%, and C-states of the 0.29\% on a real scientific workload.
In large HPC runs, COUNTDOWN leads to savings of 23.32\% on average for the NAS\cite{nas} parallel benchmarks on 1024 cores and to 22.36\% for an optimized QuantumESPRESSO (QE) on 3456 cores. When we run QE without communication tuning the savings increases to 37.74\%.
The paper is organized as follows.
Section \ref{sec:related}, presents the state-of-the-art in power and energy management approaches for scientific computing systems. Section \ref{sec:background} introduces the key concepts on power-saving in MPI phases of the application.
Section \ref{sec:framework} explains our COUNTDOWN runtime and the characterizations of real HPC applications.
Section \ref{sec:experimental} characterizes the COUNTDOWN library and report experimental results in power saving of production runs of applications in a tier1 supercomputer.

%% file: related.tex
\section{Related Work}
\label{sec:related}
% Performance penalty in favor of energy saving
%State-of-the-art works in energy-efficient HPC systems rely on DVFS mechanisms of the processing elements to reduce power consumption.
Several works focused on  mechanisms and strategies to maximize energy savings at the expense of performance. These works focus on operating the processors at a reduced frequency for the entire duration of the application \cite{fraternali_islped04,lrz_lowfreq,losalomos_sc05}. The main drawback of these approaches is the negative impact on the application performance which is detrimental to the data center cost efficiency and TCO.

Fraternali et al. \cite{fraternali_islped04,fraternali_tpds17} analyzed the impact on frequency selection on a green HPC machine which can lead a significant global energy reduction in real-life applications but can also induce significant performance penalties.
Auweter et al. \cite{lrz_lowfreq} developed an energy-aware scheduler that relies on a predictive model to predict the wall-time and the power consumption at different frequency levels for each running applications in the system.
The scheduler uses this information to select the frequency to apply to all the nodes executing the job to minimize the energy-to-solution allowing unbounded slowdown in the TTS. %system's node which represents the best operative point between performance and energy consumption for the entire system.
%TBC con che politica??? -> identifica le applicazioni tramite dei tag utenti, vengono profilate alla  frequenze massima e inserite in un database, lo scheduler ha una formulazione matematica per trovare il tempo di esecuzione di una applicazione che esegue a una determinata frequenza in base ai parametri monitorati quando è stata eseguita alla frequenza massima. Quando l'applicazione viene rieseguita viene eseguita a una frequenza per minimizzare l'ets (la frequenza magica 2.3GHz è una costatazione di LRZ in base al profiling delle loro applcazioni)
%
The main drawback of this approach is the selection of a fixed frequency for the entire application run which can cause a significant penalty on CPU-bound applications.
Hsu et al. \cite{losalomos_sc05} propose an approach where users can specify a maximum-allowed performance slowdown for their applications while the proposed power-aware runtime reduces frequency on time windows,  respecting the user's specified constraint. For this purpose, the proposed run-time estimates the instruction throughput dependency over frequency and minimizes the frequency while respecting the user's specified maximum-allowed performance slowdown. Similarly to the previous approach, energy gain is possible only by degrading the performance of the application.
%The runtime schedules CPU frequencies and voltages in such a way that the overhead does not exceed this constraint. 
%TBC such a way ??? con che politica???  b-adaptation algorith: timerizzato 1 secondo, -> alarm signal OS, ogni secondo stima un parametro B che identifica il numero di accessi off-chip e che viene utilizzato insieme a un parametro di performace loss per scegliere la frequenza dell'instante successivo, l'algoritmo si basa sul numero di istruzioni eseguite in un secondo

The main drawback of the works mentioned so far is that they lead to a systematic increase of TTS, which may be acceptable for the user, but it is not easily acceptable by the facility manager since it reduces the data center cost efficiency and TCO \cite{borghesi2018pricing}. For this reason, there is a trend in the literature towards HPC energy reduction methodologies with negligible or low impact on TTS of the running applications.

Sundriyal et al. \cite{MVAPICH2_PSTATE,MVAPICH2_ALL,MVAPICH2_Gather,MVAPICH2_TSTATE} analyze the impact of fine-grain power management strategies in MVAPICH2 communication primitives, with a focus on send/receive \cite{MVAPICH2_PSTATE}, All-to-All \cite{MVAPICH2_ALL}, and AllGather communications \cite{MVAPICH2_Gather}. In \cite{MVAPICH2_PSTATE} the authors propose an algorithm to lower the P-state of the processor during send and receive primitives. The algorithm dynamically learns the best operating points for the different send and receive calls. In the \cite{MVAPICH2_Gather, MVAPICH2_ALL, MVAPICH2_TSTATE} works, the authors propose to lower also the T-state during the send-receive, AllGather and All-to-All primitives as this increases the power savings. 
These approaches show that power saving can be achieved by entering in a low power mode during specific communication primitives but they depend on a specific MPI implementation. Differently, we show that significant savings can be achieved without impacting the implementation of the MPI library.

%In these works, authors discover that the overhead of this solution is more prominent when these communication patterns are intra-node.
%Moreover, the overhead of the proposed strategy decreases with message dimension. Authors show that this overhead is different within the different kernels implementing the communication primitives. Thus they propose a low power implementation of the studied MVAPICH2 primitives where P-state and T-state are lowered in specific stages of these primitives. These approaches show that power saving can be achieved by entering in a low power mode during specific communication primitives. These predicting algorithms to be effective need to be aware of the internal logic of the communication primitive. Thus they depend on a specific MPI implementation. Differently, we show that significant savings can be achieved without impacting the implementation of the communication primitives.

Moreover, there are other works which focus on the power consumption of network equipment during the execution of parallel applications \cite{chen2016reducing}. While these works impact on the HPC network we leverage on the power consumption of the computing units during communications.

Rountree et al. \cite{adagio_static} analyze the energy savings which can be achieved on MPI parallel applications by slowing down the frequencies of processors which are not in the critical path. Authors of the paper define tasks as the region of code between two MPI communication calls, we will refer later in the document to tasks as phases. 
The critical path is defined as the chain of the tasks which bounds the application execution time. Indeed, cores executing tasks in the critical path will be the latest ones to reach the MPI synchronization points, forcing the other cores to wait. In \cite{adagio_static} authors propose a methodology for estimating offline the minimum frequency at which the waiting cores can execute without affecting the critical path and the TTS.
In the same work, the authors suggest that the core's frequency cannot be changed too often without causing overheads. For this reason, the authors introduce a timer logic set at 10ms to avoid changing the core's frequency too often. This value is empirically found. With COUNTDOWN we demonstrate that in modern CPUs the best setting for this timer value corresponds to the built-in HW power controller latency.

%as the last process reaching the synchronization point. By mean of a linear programming model the authors of the paper estimate a reduced frequency at which the processors not in the critical path can execute \emph{without} 
%TBC è vero without???
%causing delays to the critical path processes'. 

%available energy saving of an MPI application analyzing the critical path on an offline trace.
A later work of the same authors \cite{adagio_dynamic}, implements an online algorithm to identify the 
% ANDREA MPI 
task and the minimum frequency at which it can be executed without worsening the critical path. In case the optimal frequency (which would nullify the communication blocking time) is below the minimum available one the authors propose to lower the core's frequency to the minimum one. This is done with a slack reclamation policy which is based on the measurement of the previous blocking time duration. If this was at least twice longer than an empirical time threshold (100ms) when the same task is executed again, a timer is set to the empirical threshold. If the MPI phase expires before the timer ends, nothing happens. Otherwise, when the timer expires, the core's frequency is set to the minimum one. This, in essence, implements a last-value prediction logic to determine if there will be enough blocking time which could be exploited to save energy and is only evaluated with small benchmarks (32 MPI processes). COUNTDOWN uses a timeout policy as well, but it applies it for each MPI phase without trying to predict its duration. This is a significant difference w.r.t to the \cite{adagio_dynamic} which makes it robust to miss-predictions \cite{benini_power_survey}. 
%predict the application phases to identify the critical path and consequently reduce CPU frequency when the process is not involved in the path.
Similarly, Kappiah et al. \cite{freeh2008just} developed Jitter, an online runtime based on the identification of the critical path on the application among compute nodes involved in the application run.
Liu et al. \cite{barriers_cmp} use a similar methodology as Kappiah et al. \cite{freeh2008just} but they apply it to a multi-core CPU. Zhai et al. \cite{zhai2016performance} propose a method for estimating the duration of an MPI parallel application.

The authors of \cite{simil_adagio}, as in \cite{adagio_static,adagio_dynamic}, focus on saving power by entering a low power state for processes which are not in the critical path. The authors propose an algorithm to save energy by reducing application unbalance. This is based on measuring the start and end time of each MPI\_barrier and MPI\_Allreduce primitives to compute the duration of application and MPI code. Based on that the authors propose a feedback loop to lower the P-state and T-state if in previous compute and MPI region the overhead was below a given threshold. The algorithm is based on the assumption that the duration of the current application and MPI phases will be the same as the previous ones.
In COUNTDOWN we target recent HW and larger production runs where we do not use any previous information on MPI and application phase duration, which may lead to costly performance overhead in case of misprediction in particular in irregular applications \cite{kerbyson2011energy}. Instead, COUNTDOWN relies only on a pure-reactive timer-based logic. It is worth to notice that differently from  \cite{adagio_dynamic}, the COUNTDOWN logic does not use any pre-characterization of the message-transfer time of the MPI library to estimate the communication blocking time due to this can change depending to the network congestion of the high-performance interconnect.
%which is not trivial in modern NUMA large-scale parallel systems.

To save energy during MPI phases, Lim et at. \cite{lim2006adaptive} propose to reduce core's frequency in ``long'' MPI phases. Subsequent short MPI phases are grouped and treated as a single long MPI phase. They use an algorithm to select the best P-state to be applied according to the micro-operation throughput in the MPI phase. Similarly to \cite{adagio_static,adagio_dynamic}, this approach is based on the assumption that the duration and instruction composition of current MPI phase will be the same as the previous ones. Moreover, by treating short MPI phases as a single long one, the application phases between them are executed at low frequency leading overheads.

Li et al. \cite{li2004thrifty} use a similar approach to \cite{lim2006adaptive} to reduce power consumption in synchronization points. This work focuses on collective barriers for parallel applications in shared-memory multiprocessors. Differently, from the previous approaches, instead of using P-state, they use idle states (C-states) and specific hardware extensions to account for their transitioning (sleep and wake-up) times. As in the previously described approaches, this runtime uses a history-based prediction model to identify the duration of the next barriers.

\begin{figure*}
    \centering
    \captionsetup[subfigure]{position=top}
    \subfloat[All MPI processes are involved in the diagonalization QE-CP-EU]{\includegraphics[scale=0.358]{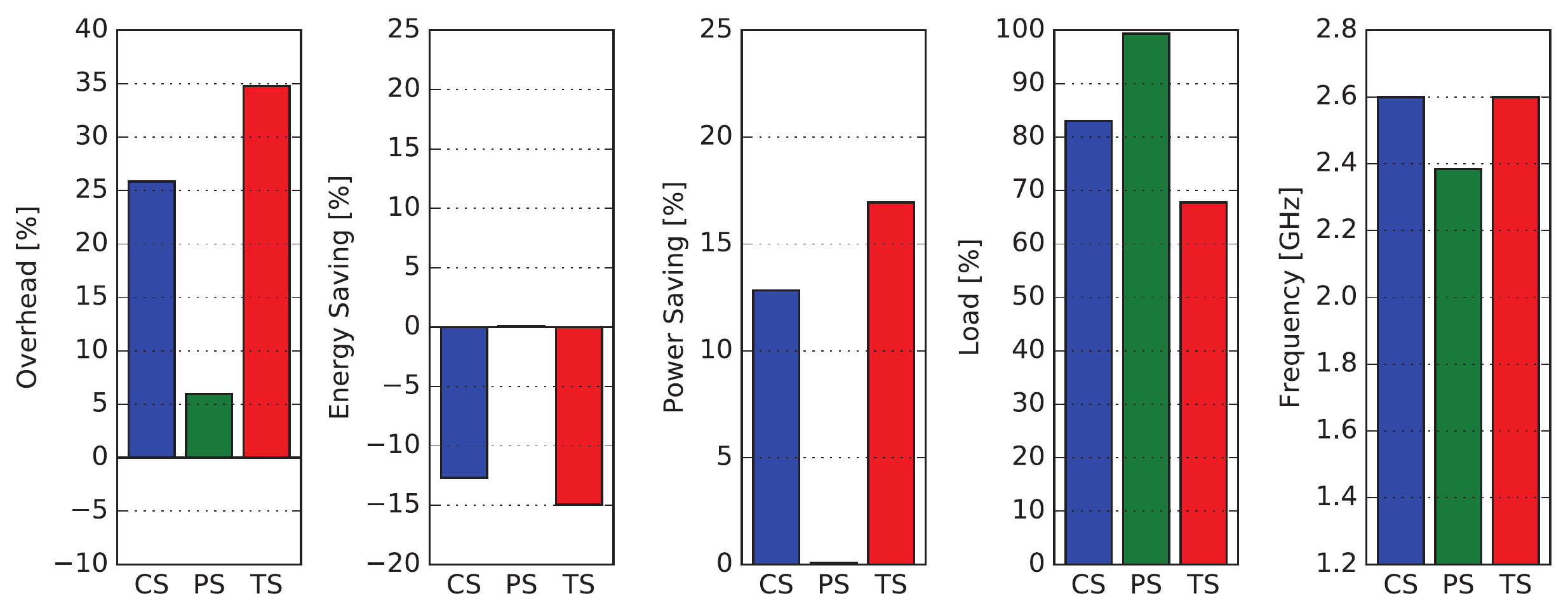} }
    \subfloat[Single MPI process is involved in the diagonalization QE-CP-NEU]{\includegraphics[scale=0.358]{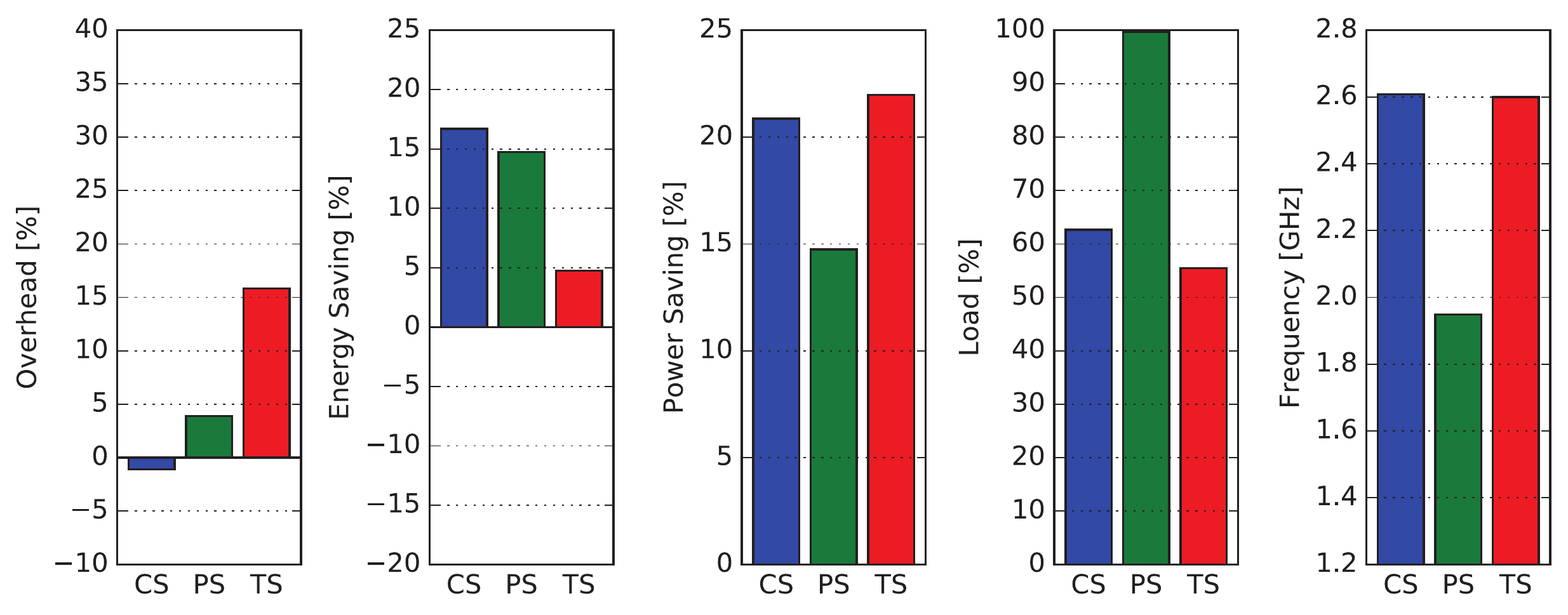} }
    \caption{Overhead, energy/power saving, average load and frequency for QE-CP-EU (a) and QE-CP-NEU (b). Legend: C-state ($CS$), P-state ($PS$) and T-state ($TS$) mode. Baseline is busy-waiting mode (default mode) of MPI library.}
    \label{fig:cs_ps_ts}
\end{figure*}

The authors of \cite{MVAPICH2-EA} show that the approaches in \cite{adagio_static,adagio_dynamic} and the ones which estimate the duration of MPI and communication phases based on a last-value prediction\cite{lim2006adaptive,li2004thrifty} can lead to significant misprediction errors. The authors propose to solve this issue by estimating the duration of the MPI phases with a combination of communication models and empirical observation specialized for the different groups of communication primitives. If this estimated time is long enough, they will reduce the P-state. As we will show with the proposed COUNTDOWN approach, this can be achieved without a specific library implementation and communication models.

Li et al. \cite{Schulz_IPDPS10} analyzed hybrid MPI/OpenMP applications in term of performance and energy saving and developed a power-aware runtime that relies on dynamic concurrency throttling (DCT) and DVFS mechanisms.
This runtime uses a combination of a power model and a time predictor for OpenMP phases to select the best cores' frequency when application manifests workload imbalance.

The works in the second group, namely \cite{freeh2008just,barriers_cmp,Schulz_IPDPS10,lim2006adaptive,li2004thrifty}, but also \cite{adagio_dynamic} in the slack reclamation policy, have in common the prediction of future workload imbalances or MPI phases obtained by analyzing previous communication patterns. However, this approach can lead to frequently mispredictions in irregular applications \cite{kerbyson2011energy} which cause performance penalties. COUNTDOWN differs from the above approaches (and complements them) because it is purely reactive and does not rely on assumptions and estimation of the future workload unbalance.

The power management literature has analyzed in depth the issue of prediction inaccuracy and predictive model overfitting \cite{benini_power_survey}. One of the key outcome of COUNTDOWN, is that timeout-based policies are effective if predictions are not available (e.g. when data is being collected for building a predictive model), and are also essential in mitigating miss prediction overheads. 
%Timeout-based power management policies are simple to calibrate, as they only require to find a timeout threshold that prevents power state changes when the duration of the slow-down (or shut-down) phase is not long enough to amortize power-state transition overhead. In the following Sections, we will show how to identify the critical timeout threshold in the specific context of MPI applications running on HPC machines. 

%Conficoni et al. \cite{CONFICONI_DATE15} show that HPC cooling costs are mostly related to a large number of factors, on the most relevant: the total power consumption, ambient temperature and cooling control policy.
%The temperature can change fast showing high difference between nodes and CPUs.
%As a side effect, energy saving strategies ease the thermal control by reducing power consumption and core's temperature.
%Energy saving strategies also reflect on thermal control for supercomputing reducing overheating situations.

The implementation of power management strategies greatly benefits from standard APIs and platform-independent software abstractions to interface with the hardware. Eastep et al. propose GEOPM \cite{GEOPM}, an extensible and plug-in based framework for power management in large parallel systems.
GEOPM is an open-source project and exposes a set of APIs that programmers can insert into applications to combine power management strategies and HPC workload.
A plugin of the framework targets power constraint systems aiming to speed up the critical path migrating power to the CPU's executing the critical path tasks.
In a similar manner, another plugin can selectively reduce the frequency of the processors in specific regions of codes flagged by the user by differentiating regions in CPU, memory, IO, or disk bound.
%TBC: GEOPM per funzionare ha bisogno o no di cambiare il codice dell'applicazione??
% anche loro usano PMPI e openmp, per fare cosa? GEOPM può usare una instrumentazione del codice utente ma questa non è obbligatorio. è chiaro che se il codice utente viene istrumentato il profiling dell'applicazione e del tipo di fasi (CPU vs memory bound) è migliore, quindi GEOPM può applicare delle politiche con una grana più fine e risparmiare più energia o aumentare le performance.
Today, GEOPM is capable of identifying MPI regions and reducing the frequency based on MPI primitive type.
However, it cannot differentiate between short and long MPI and thus cannot control the overhead caused by the frequency changes and runtime in short MPI primitives. COUNTDOWN addresses this limitation and can be integrated into future releases of GEOPM, as its design principles are entirely compatible with it (i.e. no application code modifications are required).

%previous version
An earlier version of the COUNTDOWN run-time was presented in \cite{ANDARE}. This paper adds in COUNTDOWN the support for two additional low power state mechanisms (C-state and T-states) and their comparisons with P-state. Moreover, we extended \cite{ANDARE} with a detailed analysis of the timeout configuration for the three different low power state mechanisms, and the implication of the timeout with the MPI and application phases duration. We finally extended \cite{ANDARE} with a broader set of experimental results, including the NAS parallel benchmarks and an additional QE large-scale run with different network optimization, which is a common use-case in supercomputer environment.

%% file: background.tex
\section{Background}
\label{sec:background}

%ANDREA
In this section, we show the implications and challenges of transitioning into low power states (P/C/T-states) during synchronization and communication primitives for energy-savings on two practical examples.

%As previously highlighted today's power management approaches for HPC systems lack the support for taking advantages of the slack induced by synchronization and communication primitives. Indeed, as described in previous section processors HPC systems have the capability of changing the performance and power consumption trade-off dynamically by entering in idle (shutdown) and active (DVFS) low power states, thus in practice, there is no limitation for taking advantage of them during MPI communication and synchronization phases.
%Power management strategies for HPC applications often leverage on MPI primitives to reduce energy consumption.
%The main limitation is the performance penalty imposed on the application caused by platform constraints.
%In this section we give two practical examples on the drawbacks and limitation of doing it. 
%approach in a single compute node.  
%We can recognize two families of approaches: (i) bringing the core in idle low power mode or (ii) in an active low power mode each time the execution encounters an MPI call.
%We realize a set of experiments on a single-compute node and on an HPC system to show the impact on both performance and energy consumption.
%% ANDREA non capisco la frase seguente.
%In the next Sections, we will show single-node exploration using state-of-the-art power management strategies for MPI libraries to reduce energy and power consumption leveraging on the MPI primitives.

%\subsection{Target Architecture and Benchmark}

As a test platform we have used a compute node equipped with two Intel Haswell E5-2630 v3 CPUs, with 8 cores at 2.4 GHz nominal clock speed and 85W Thermal Design Power (TDP) and 
%% DANIELE MOVE THIS PART IN EXPERIMENTAL RESULTS: 
%Instead, for multi-node experiments we use a Tier-0 HPC system based on an IBM NeXtScale cluster which is currently classified in the Top500 supercomputer list \cite{TOP500}.
%The compute nodes of the HPC system, are equipped with 2 Intel Broadwell E5-2697 v4 CPUs, with 18 cores at 2.3 GHz nominal clock speed and 135W TDP and interconnected with an Intel QDR (40Gb/s) Infiniband high-performance network.
the production software stack of Intel systems\footnote{We use Intel \textit{MPI Library 5.1} as the runtime for communication, coupled with Intel \textit{ICC/IFORT 18.0} as our toolchain. We choose Intel software stack because it is currently used in our target systems as well supported in most of HPC machines based on Intel architectures}.
We use a single compute node for the following exploration because this is a worst-case scenario for energy-saving strategies in MPI applications due the communications happen in a very short time.

All the tests in this Section have been executed on a real scientific application, namely QuantumESPRESSO which is a suite of packages for performing Density Functional Theory based simulations at the nanoscale and it is widely employed to estimate ground state and excited state properties of materials \emph{ab initio}. For these single nodes tests, we used the CP package parallelized with MPI. We use QE because it is a paradigmatic application that shows the typical behaviors of HPC codes. QE main computational kernels include dense parallel linear algebra (diagonalization) and 3D parallel FFT, which makes the following exploration work relevant for many HPC codes\footnote{QE mostly used packages are: (i) \textit{Car-Parrinello} (CP) simulation, which prepares an initial configuration of a thermally disordered crystal of chemical elements by randomly displacing the atoms from their ideal crystalline positions; (ii) PWscf (Plane-Wave Self-Consistent Field) which solves the self-consistent Kohn and Sham (KS) equations and obtain the ground state electronic density for a representative case study \cite{avvisati2017fepc}}.

%In the following we will only refer to data obtained with pure MPI parallelism since this is the main focus of this article and this choice does not significantly impair the conclusions reported later
%The code uses a pseudo-potential and plane-wave approach and implements multiple hierarchical levels of parallelism performed with a hybrid MPI+OpenMP approach. As of today, OpenMP is generally used when MPI parallelism saturates, and it can improve the scalability in the highly parallel regime. Nonetheless,

To exploit the system behavior for different workload distribution in a single node evaluation, we focused the computation of the band structure of the Silicon along the main symmetry. When executed by a user with no domain expertise and with default parameter, QE runs with a hybrid MPI parallelization strategy with only one MPI process used to perform the diagonalization and all the MPI processes used to perform the FFT kernel. We will later refer to this case as \textit{QuantumESPRESSO CP Not Expert User} (QE-CP-NEU). Differently, when an expert user runs the same problem, he changes the parameters to better balance the workload by using multiple MPI processes to parallelize also the diagonalization kernel. We will later refer to this case as \textit{QuantumESPRESSO CP Expert User} (QE-CP-EU).
In the QE-CP-NEU case, when a single process works on the linear algebra kernel, the other ones remain in busy wait on the MPI call. In the following text, we will compare fine-grain power management solutions with the busy-waiting mode (default mode) of MPI library, where processes continuously poll the CPU for the whole waiting time in MPI synchronization points.

%We take into exam the case of the diagonalization workload on a single process because we will show how this configuration is impacted by low power states and HW power controller-based boosting.

\begin{figure}[t]
    \centering
    \includegraphics[width=1.00 \linewidth]{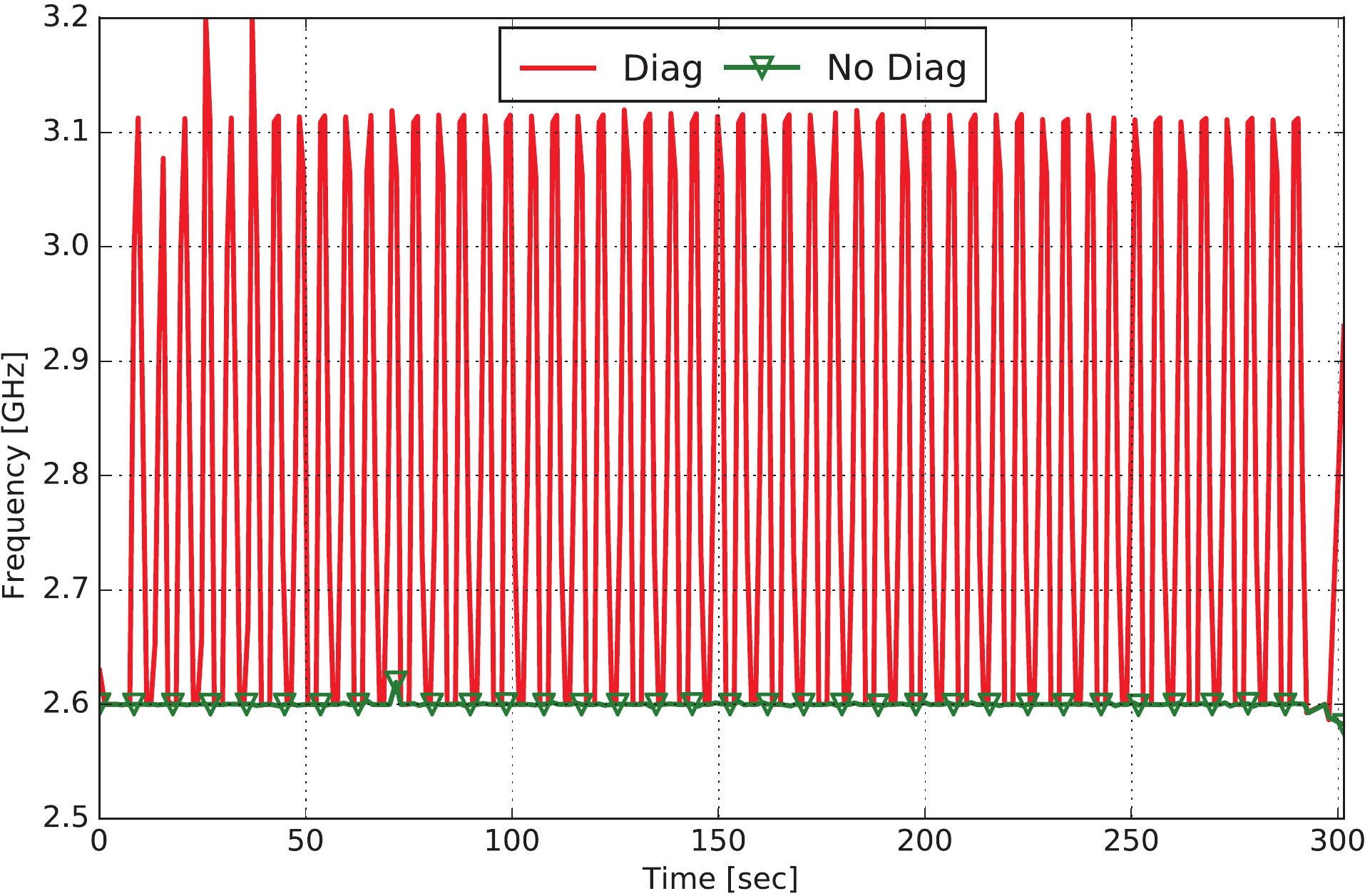}
    \caption{Time plot of frequency for QE-CP-NEU identifies the frequency of the MPI process working on the diagonalization, \textit{No Diag} is the average frequency of MPI processes not involved in the diagonalization.}
    \label{fig:turbo}
\end{figure}

\subsection{Wait-mode/C-state MPI library}
\label{sec:wait_mode}
Usually, MPI libraries use a busy-waiting policy in collective synchronizations to avoid performance penalties.
This is also the default behavior of Intel MPI library.
This library can also be configured to release the control to the idle task of the operating system (OS) during waiting time to leverage the C-states of the system.
This allows cores to enter in sleep states and being woken up by the MPI library when the message is ready through an interrupt routine.
In the Intel MPI library, it is possible to configure the wait-mode mechanism through the environment variable \textit{I\_MPI\_WAIT\_MODE}.
This allows the library to leave the control to the idle task, reducing the power consumption for the core waiting in the MPI.
The transitions in and out from the sleep mode induce overheads in the execution time.

In figure \ref{fig:cs_ps_ts} are reported the experimental results, the wait-mode strategy is identify with \textit{CS}.
From it, we can see the overhead induced by the wait mode w.r.t. the default busy-waiting configuration, which worsens by 25.85\% the execution time.
This is explained by the high number of MPI calls in the QE application which leads to frequent sleep/wake-up transitions and high overheads.
From the same figure, we can also see that the energy saving is negative, which is -12.72\%, this is because the power savings obtained in the MPI primitives does not compensate the overhead induced by the sleep/wake-up transitions.
Indeed, the power reduction is of 12.83\%. This is confirmed by the average load of the system, which is 83.02\% as the effect of the C-states activity in the MPI primitives. The average frequency is 2.6GHz, which is the standard turbo frequency of our target system.

Surprisingly, the QE-CP-NEU case has a negative overhead (-1.08\%\ overhead is a speedup).
This speedup is given by the turbo logic of our system.
Indeed, we can see that the average frequency is slightly higher than 2.6GHz, which means that the process doing the diagonalization can leverage the power budget freed by the other processes not involved in the diagonalization while they are waiting in a sleep state in the MPI runtime.
In figure \ref{fig:turbo}, we report the average frequency of the process working on the diagonalization and the average frequencies of all the other MPI processes.
In the target system, a single core can reach up to 3.2 GHz if only one core is running, this is what happens when all cores are waiting in a sleep state for the termination of the diagonalization workload.
The benefit of this frequency boosting unleashed by the idle mode on the MPI library and the unbalanced workload can save up to 16.69\% of energy with a power saving of 20.86\%.

As a conclusion of this first exploration, we recognize that it is possible to leverage the wait mode of the MPI library to save power without increasing the execution time, but energy savings and impact on the TTS depends on the MPI calls granularity which can lead to significant penalties if the application is characterized by frequent MPI calls.

\begin{figure*}[t]
    \centering
    \includegraphics[width=1.00 \linewidth]{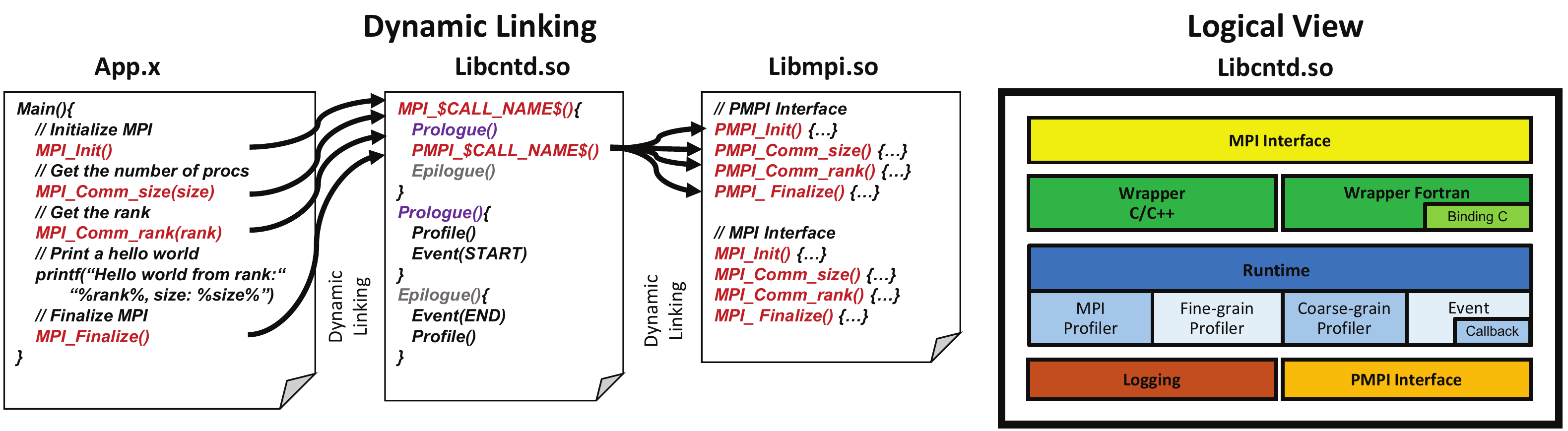}
    \caption{Dynamic linking events when COUNTDOWN is injected at loading time in the application and logical view of all the components.}
    \label{fig:dynlink_logview}
\end{figure*}

\subsection{DVFS/P-state MPI library}
To overcome the overheads of C-state transitions, we focus our initial exploration on the active low power states (C-state) and DVFS (P-state).
Intel MPI library does not implement such a feature, so we manually instrumented all the MPI calls of the application with a \textit{epilogue} and \textit{prologue} function to scale down and raise up the frequency when the execution enters and exits from an MPI call.
To avoid interference with the power governor of the operating system, we disabled it in our compute node granting the complete control of the frequency scaling.
We use the MSR driver to change the current P-state writing \textit{IA32\_PERF\_CTL} register with the highest and lowest available P-state of the CPU, which corresponds to the turbo and 1.2GHz operating points.
%This instrumentation emulates the behavior of the idle mode of Intel MPI library but with P-states instead of C-states and without interrupt routines.
In figure \ref{fig:cs_ps_ts} we report the results of this exploration, where the P-state case is labelled with \textit{PS}.

%\begin{figure*}[t]
%    \centering
%    \includegraphics[width=1.00 \linewidth]{./images/dynamic_linking}
%    \caption{Dynamic linking events when COUNTDOWN is injected at loading time in the application.}
%    \label{fig:dynamic_linking}
%\end{figure*}

In the overhead plot, in figure \ref{fig:cs_ps_ts}.a, we can see that the overhead is significantly reduced w.r.t C-state mode, reducing the 25.85\% overhead obtained previously to 5.96\%.
This means that the overhead of scaling the frequency is lower respect to the sleep/wake-up transitions cost.
However, the energy and power savings are almost zero. Similarly to QE-CP-EU, this happens because all the MPI processes participate in the diagonalization, thus we have a high number of MPI calls with very short duration.
This is also confirmed by the average frequency, which does not show significant variations w.r.t. the busy waiting, with a measured average frequency of 2.4GHz.
The load bar reports 100\% of activity, which means that there is no idle time as expected.

Focusing in the QE-CP-NEU case, in figure \ref{fig:cs_ps_ts}.b, the overhead is 3.88\% which is reduced w.r.t. QE-CP-EU. In addition, in this case, we have significant energy and power saving, respectively of 14.74\% and 14.75\%.
These saving are due to the workload unbalance and to the long time spent in the MPI calls from the processes not involved in the diagonalization.
This is confirmed by the lower average frequency (1.95GHz). The load is unaltered as expected.

In conclusion, using DVFS for fine-grain power management instead of the idle mode allows to control the overhead for both balanced and unbalanced workload better.
However, the overhead is still significant and in HPC the TTS is the prime goal.
%Moreover, the energy saving is also dependent on the execution time, so high overhead reduces the achievable energy efficiency.
%In the next Section we will propose a study and a new methodology namely COUNTDOWN, to control the overhead induced by the fine-grain power management for MPI application. 

%\begin{figure}[t]
%    \centering
%    \includegraphics[width=1.00 \linewidth]{./images/framework}
%    \caption{Logical view of COUNTDOWN components.}
%    \label{fig:framework}
%\end{figure}

\subsection{DDCM/T-state MPI library}
One crucial question is: are the overheads of fine-grain power management strategies induced by the specific power management states? To answer this question, we considered duty-cycling low power states\footnote{In this Section we also tried to use the Dynamic Duty Cycle Modulation (DDCM) (also known as throttling states or T-states) available in the Intel architectures which are characterized by lower overhead.
DDCM has been supported in Intel processors since Pentium 4 and enables on-demand software-controlled clock modulation duty cycle.}.
In Intel CPUs, DDCM is used by the \textit{HW power controller} to reduce the power consumption when the CPU identifies thermal hazards.
%DDCM is used to throttling the CPU to reduce the temperature.
%We want to use DDCM as an alternative of DVFS to reduce the power draw in the HPC applications.
Similarly to \cite{DDCM}, we use DDCM to reduce the power consumption of the cores in MPI calls.
We manually instrumented the target as we did in the \textit{prologue} function of each MPI call to configuring DDCM to 12.5\% of clock cycles, which means for each clock cycle we gate the next 7; while in the \textit{epilogue} function, we restore the DDCM to 100\% of clock cycles, we control it by writing to the DDCM configuration register, called \textit{IA32\_CLOCK\_MODULATION}, through the MSR driver.

In figure \ref{fig:cs_ps_ts}.a, the DDCM results are reported with \textit{TS} bars.
Surprisingly, the overheads induced by T-states are greater than the wait mode and equal to 34.78\%.
As a consequence, the energy saving is the worst, leading to an energy penalty of 14.94\%.
The load is significantly reduced owing to the throttling, in an average of 67.78\%, while the frequency is constant to 2.6GHz.

In figure \ref{fig:cs_ps_ts}.b, we report T-state results for QE-CP-NEU.
Even for this unbalanced workload case, the T-states are the worst.
T-state transitions introduce an overhead of 15.82\% consequent of the power reduction, with a very small energy saving, only of the 4.75\%, and a power saving of 21.97\%.
The load of the system is reduced to 55.45\%, similarly to the idle mode, and the frequency remained unchanged as expected.

As a matter of fact, we show that phase agnostic fine-grain power management leads to significant application overheads which may nullify the overall saving. Though, we need to bring knowledge of the workload distribution and the communication granularity of the application in the fine-grain power management.
In the next Sections, we introduce the COUNTDOWN approach which addresses this issue.

\begin{figure*}
    \centering
    \includegraphics[width=1.00 \linewidth]{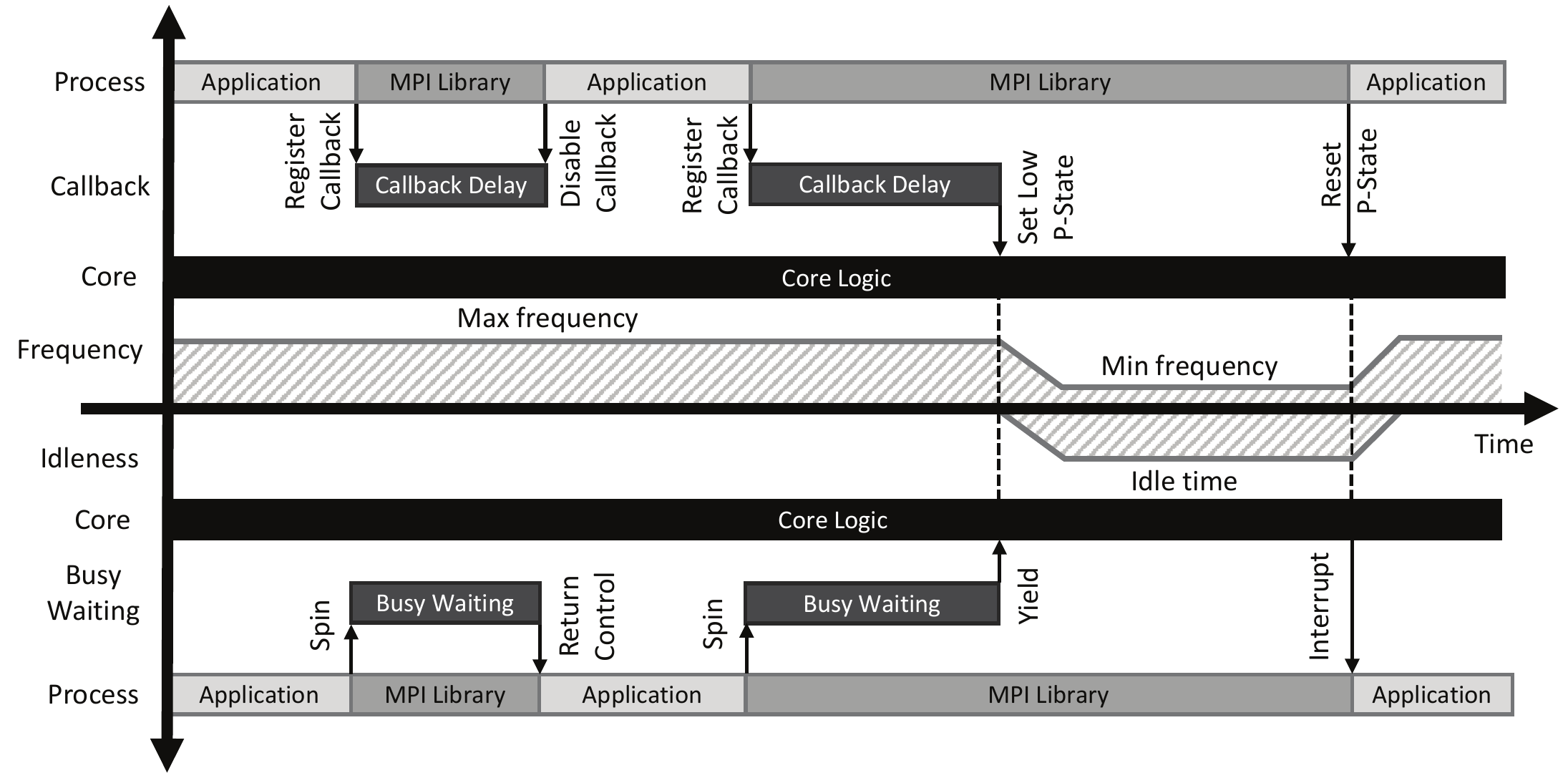}
    \caption{On the upper side is depicted the timer strategy utilized in COUNTDOWN, while in the lower side is depicted the idle-wait mode with timer implemented in the Intel MPI library.}
    \label{fig:callback}
\end{figure*}

%% file: framework.tex
\section{Framework}
\label{sec:framework}

%We address the lacks in today fine-grain power management solutions in HPC highlighted in the previous Section by proposing COUNTDOWN.
COUNTDOWN is a run-time library for profiling and fine-grain power management written in C language.
COUNTDOWN is based on a \textit{profiler} and on a \textit{event} module to inspect and react to MPI primitives.
The key idea in COUNTDOWN can be summarized as follows. Every time the application calls an MPI primitive, COUNTDOWN intercepts the call with minimal overhead and uses a timeout strategy \cite{benini_power_survey} to avoid changing the power state of the cores during fast application and MPI context switches, where doing so may result only in state transition overhead without significant energy and power reduction. 
%In a nutshell, each time the MPI library asks to enter in low power mode, COUNTDOWN defers the decision for a defined amount of time (the timeout). If the MPI phase terminates within this amount of time, the hardware does not enter in low power state. Thus, COUNTDOWN filters short MPI phases with no potential for power-management-based energy reduction.

%The timer strategy will automatically skip short MPI calls applying our power policy only in long MPI calls.
%This is also supported in \cite{benini_power_survey} where is shown that timer strategies outperform prediction of 

In figure \ref{fig:dynlink_logview} the COUNTDOWN's components are depicted.
COUNTDOWN exposes the same interface as a standard MPI library and intercepts all MPI calls from the application.
COUNTDOWN implements two wrappers to intercept MPI calls: i) the first wrapper is used for C/C++ MPI libraries, ii) the second one is used for FORTRAN MPI libraries.
This is mandatory since C/C++ and FORTRAN MPI libraries produce assembly symbols which are not application binary (ABI) compatible.
%DANIELE quali two wrappers??? non capisco di quali stai parlando.
The FORTRAN wrapper implements (un)marshalling interfaces to bind MPI FORTRAN handlers into compatible MPI C/C++ handlers.

When an application is instrumented with COUNTDOWN, every MPI call is enclosed in a corresponding wrapper function that implements the same signature.
The wrapper function calls the equivalent PMPI call, but after and before a \textit{prologue} and an \textit{epilogue} routine.
Both routines are used by the profile and by the event modules to support monitoring and power management, respectively. 
COUNTDOWN interacts with the \textit{HW power manager} through a specific \textit{Events} module in the library.
The \textit{Events} module can also be triggered by system signals registered as callbacks for timing purposes.
COUNTDOWN configurations can be done through environment variables, and it is possible to change the verbosity of logging and the type of HW performance counters to monitor.

The library targets the instrumentation of applications through dynamic linking, as depicted in figure \ref{fig:dynlink_logview}, without user intervention.
When dynamic linking is not possible COUNTDOWN has also a fallback, a static-linking library, which can be used while building the application, to add COUNTDOWN at compilation time.
The advantage of using the dynamic linking is the possibility to instrument every MPI-based applications without any modifications of the source code nor the toolchain, even without re-compiling it.
Linking COUNTDOWN to the application is straightforward: it is enough to configure the environment variable \textit{LD\_PRELOAD} with the path of COUNTDOWN library and lunch the application as usual.

%\textbf{DANIELE} \textit{hai dato 30 mila dettagli senza dire cosa fa COUNTDOWN. Qui devi dire che:
%1. Countdown risolve i problemi evidenziati nella sezione precedente.
%2. Come lo fa:
%2a. lo fa definendo un profiler
%3a. creando due routine una di prologo ed una di epilogo atte ad entrare in low power mode solo durante le fasi mpi lunghe.
%4a qui introduci la logica. Ovvero tramite una callback con timer settato a 500us. Come ci spiega vecchio paper di luca usare un principio di località temporale funziona meglio che usare un predittore della durata futura della fase MPI. Da riscrivere tutte le righe sopra.}

\subsection{Profiler Module}
COUNTDOWN allows extracting traces, which can be exploited to estimate application performance as \cite{zhai2016performance}. COUNTDOWN uses three different profiling strategies targeting different monitoring granularity.

(i) The \textit{MPI profiler} is responsible for collecting all information regarding the MPI activity.
For each MPI process, it collects information on MPI communicators, MPI groups and the coreId.
In addition, the COUNTDOWN run-time library profiles each MPI call by collecting information on the type of the call, the entrance and exit times and the data exchanged with the other MPI processes.

(ii) The \textit{fine-grain micro-architectural profiler}, collects micro-architectural information at every MPI call along with the \textit{MPI profiler}.
This profiler uses the user-space RDPMC instruction to access the performance monitoring units implemented in Intel's processors.
It monitors the average frequency, the time stamp counter (TSC) and the instructions retired for each MPI call and application phase. It can access up to 8 configurable performance counters that can be used to monitor user-specific micro-architectural metrics.

(iii) The \textit{coarse-grain profiler} monitors a larger set of HW performance counters available in the Intel architectures.
In Intel architectures, privileged permissions are required to access HW performance counters. Such level of permissions cannot be granted to the final users in production machines.
To overcome this limitation, we use the MSR\_SAFE \cite{MSR_SAFE} driver% to access to the model-specific registers of the system (MSR)
, which can be configured to grant access to standard users on a subset of privileged architecture registers, while avoiding security issues.
At the core level, COUNTDOWN monitors TSC, instructions retired, average frequency, C-state residencies, and temperature. At uncore level, it monitors CPU package energy consumption, C-state residencies, and temperature of the packages.
This profiler uses Intel Running Average Power Limit (RAPL) to extract energy/power information from the CPU.
The \textit{coarse-grain profiler}, due to the high overhead needed by every single access to the set of HW performance counters monitored, uses a time-based sample rate.
Data are collected at least Ts second delay from the previous collection.
The \textit{fine-grain micro-architectural profiler} at every MPI calls checks the time stamp of the previous sample of \textit{coarse-grain profiler} and, if it is above Ts seconds, triggers it to get a new sample.
These capabilities are added to the application through the \textit{prologue} and \textit{epilogue} functions as shown in figure \ref{fig:dynlink_logview}.

COUNTDOWN also implement a logging module to store profile information in a text file which can be written in local or remote storage.
While the log file of MPI profiler can grow with the number of MPI primitives and can become significant in long computation (thus the information is stored in binary files), the logging module also reports a summary of this information in an additional text file. 

\begin{figure*}[t]
    \centering
    \includegraphics[width=1.00 \linewidth]{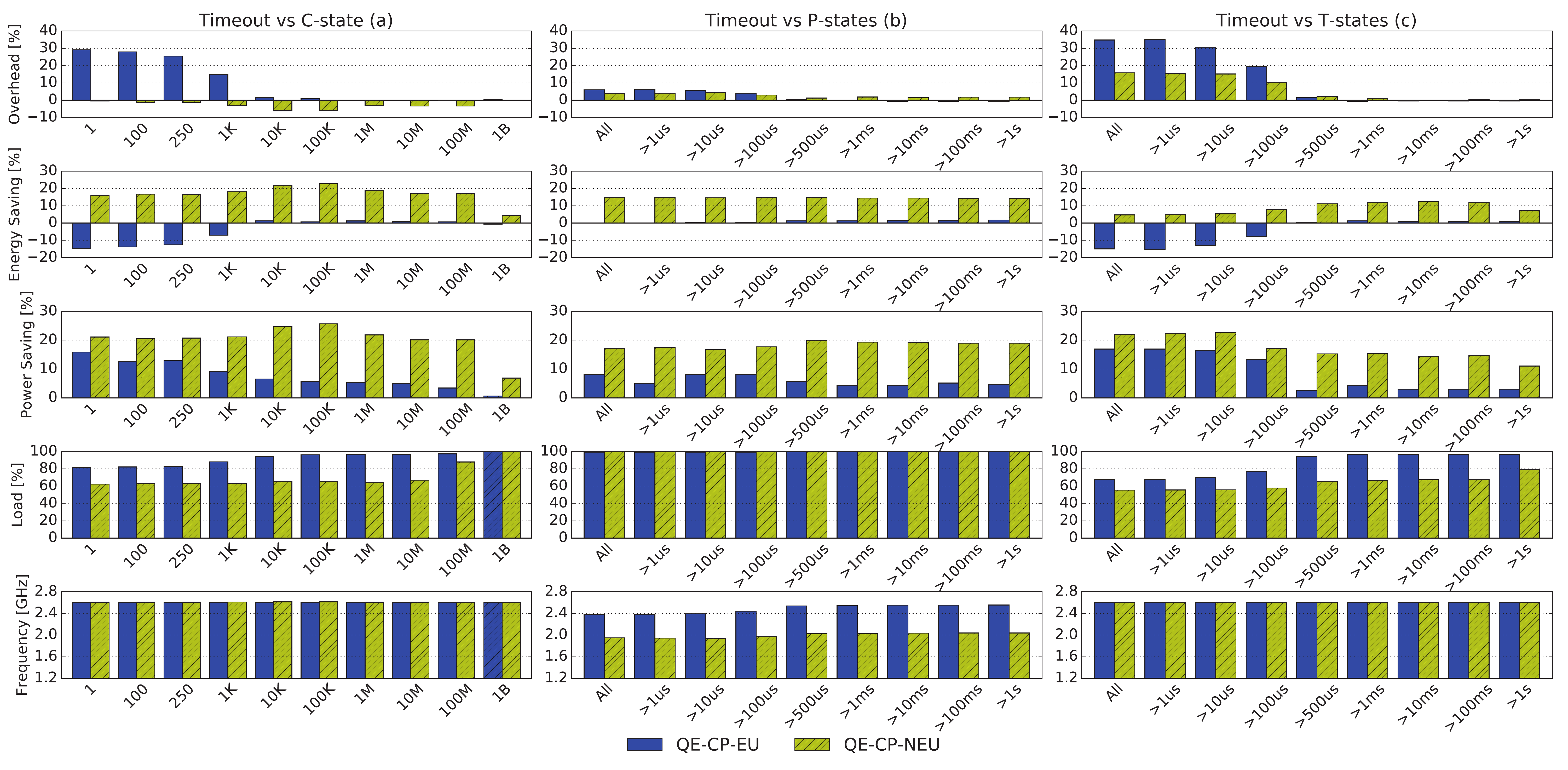}
    \caption{Impact of MPI phases duration on the overheads, energy and power savings of C/P/T-state for QE-EU and QE-NEU.}
    \label{fig:timeout_cases}
\end{figure*}

\subsection{Event Module}
\label{sec:event_module}
COUNTDOWN interacts with the \textit{HW power controller} of each core to reduce the power consumption.
It uses MSR\_SAFE to write the architectural register to change the current P-state independently per core.
When COUNTDOWN is enabled, the \textit{Events} module select the performance level at which to execute a given phase.
%We also extended this functionality with a callback system to.
%\textbf{DANIELE non chiaro perchè ti serve la callback. Serve dire ci serviva questa e questa funzionalità e l'abbiamo messa tramite una call back.}

%DANIELE definisci qui il prologo e l'epilogo di countdown. -> ho definito sopra le funzioni di prologo ed epilogo perchè vengono anche utilizzate nella fasi di profiling. Qui aggiungo che sempre in queste funzioni viene registrato il timer con la callback
COUNTDOWN implements a timeout strategy through the standard Linux timer APIs, which expose the system calls: \textit{setitimer()} and \textit{getitimer()} to manipulate user-space timers and register callback functions.
This methodology is depicted in figure \ref{fig:callback} in the top part.
When COUNTDOWN encounters an MPI phase, in which opportunistically can save energy by entering in a low power state, \textit{registers} a timer callback in the \textit{prologue} function (Event(start)), after that the execution continues with the standard workflow of the MPI phase.
When the timer expires, a system signal is raised, the ``normal'' execution of the MPI code is interrupted, the signal handler triggers the COUNTDOWN \textit{callback}, and once the callback returns, execution of MPI code is resumed at the point it was interrupted.
If the ``normal'' execution returns to COUNTDOWN (termination of the MPI phase) before the timer expiration, COUNTDOWN \textit{disables} the timer in the \textit{epilogue} function and the execution continues like nothing happened.
The callback can be configured to enter in the lower T-state (12.5\% of load), later referred to as \textit{COUNTDOWN THROTTLING}, or in the lower P-state (1.2GHz) later referred to as \textit{COUNTDOWN DVFS}.

Intel MPI library implements a similar strategy, but it relies on the sleep power states of the cores. Its behavior is depicted in the bottom part of figure \ref{fig:callback}.
If the environment variable \textit{I\_MPI\_WAIT\_MODE}, presented in Section \ref{sec:wait_mode}, is combined with the environment variable \textit{I\_MPI\_SPIN\_COUNT}, it is possible to configure the spin count time for each MPI call. When the spin count becomes zero, the MPI library leaves the execution to the idle task of the CPU.
This parameter does not contain a real-time value but includes a value which is decremented by the spinning procedure on the MPI library until it reaches zero.
This allows the Intel MPI library to spin on a synchronization point for a while, and after that, enter in an idle low power state to reduce the power consumption of the core.
The execution is restored when a system interrupt wakes up the MPI library signaling the end of the MPI call. Later, we will refer to this mode as \textit{MPI SPIN WAIT}.

In the next Section, we will clarify though experiment why the timeout logic introduced by COUNTDOWN is effective in making fine-grain power management possible and convenient in MPI parallel applications.

%% file: experimental.tex
\section{Experimental Results}
\label{sec:experimental}
In this Section, we present: (i) an overhead analysis of COUNTDOWN, (ii) the effect of timeout strategy using different timeout delays, and (iii) the evaluation on a single node and a production HPC system with real scientific applications.

\subsection{Framework Overheads}
%%DANIELE
%%\textbf{DANIELE} \tetit{Questo è il test con l'applicazione sempre della frequenza massima o quello puro profiler? Hai il tempo di quello con il cambio della frequenza massima sempre? Sarebbe bene riportarlo. Spiegalo meglio. -> fatto}

We evaluate the overhead of running MPI applications instrumented with the profiler module of COUNTDOWN without changing the cores' frequency.
We run QE-CP-EU on a single node, which is the worst case for COUNTDOWN in term of number and granularity of MPI calls to profile because all network-related overheads in MPI calls are nullified and intra-chip communication and synchronization are orders-of-magnitude faster than the inter-chip or inter-node ones. Hence, MPI wait-times exploitable for power management are generally much shorter. 
%Indeed, this particular QE configuration is quite stressful for the profiling module of COUNTDOWN.

In this run, there are more than 1.1 million of MPI primitives for each process in the diagonalization task: our run-time library needs to profile in average an MPI call every 200us for each process.
We measured the overhead comparing the execution time with and without COUNTDOWN instrumentation. We repeated the test five times, and we report the median case.
Our results show that even in this unfavorable setting, the COUNTDOWN profiler introduces an overhead in the execution time which is less than 1\%.
%This result is also supported by the overhead measurements of the prologue and the epilogue routines that COUNTDOWN injects in the application for each MPI call, which costs between 1us and 2us.
%We remark that these results show that the instrumentation of all the MPI calls per se is not a source of overhead.
We repeated the same test changing the cores' frequency to assess the overhead of a fine-grain DVFS control.
To measure only the overhead caused by the interaction with the DVFS knobs, we force COUNTDOWN to force always the highest P-state in the DVFS control registers.
Thus, we avoid application slowdowns caused by frequency variation, and we obtained only the overhead caused by the register access.
Our experimental results report of 1.04\% of overhead to access the DVFS control register and for the profile routines.

These results prove that the source of the overheads of phase agnostic fine-grain power management is not related to issuing the low power state transition (DVFS in this case).
%Thus, why results in Section \ref{sec:background} had these large overheads?
Figure \ref{fig:timeout_cases} focuses on understanding the source of this by replicating the tests of Section \ref{sec:background} for both QE-CP-EU and QE-CP-NEU, but now entering in the low power state only for MPI phases longer than a given time threshold. For the P-state and T-state (Figure \ref{fig:timeout_cases}.b and Figure \ref{fig:timeout_cases}.c) we obtained that by profiling in advance the duration of each MPI phase and instrumenting with the low power command only the phases which had a duration longer than the threshold. We report on the x-axes the time threshold value. For C-state (Figure \ref{fig:timeout_cases}.a) we leveraged the COUNTDOWN MPI logic, \textit{I\_MPI\_SPIN\_COUNT} parameter to filter out short phases. On the x-axis, we report the \textit{I\_MPI\_SPIN\_COUNT} parameter.

From the plot, we can recognize that there is a well-defined threshold of 500us for the T-state and P-state case and of 10K iteration steps for the C-state after which the overhead introduced by the fine-grain power management policy is reduced and the energy savings becomes positive for the QE-CP-EU. In the next Section, we will analyze why this happens by focusing on the P-state case.

The overhead in term of memory is negligible since the memory required by COUNTDOWN is just a few megabytes for each MPI process.

\begin{figure*}[t]
    \centering
    \includegraphics[width=1.00 \linewidth]{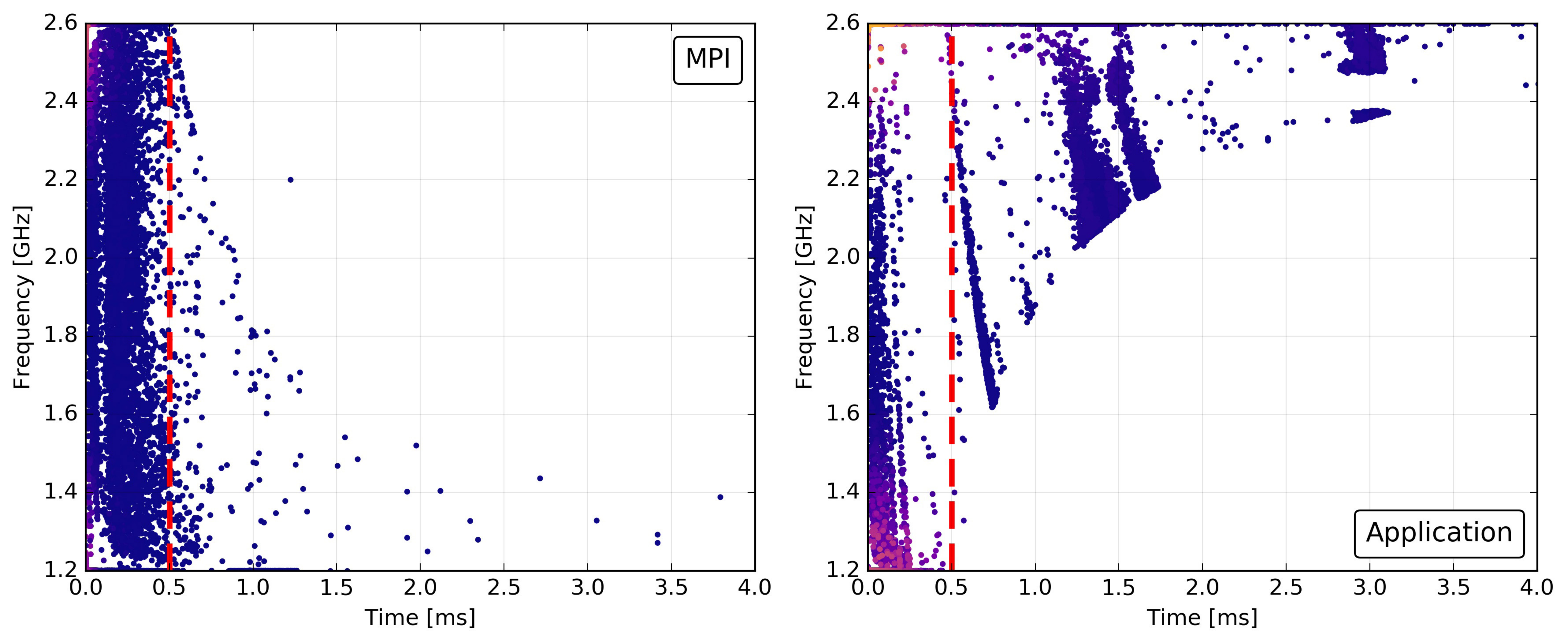}
    \caption{Average frequency and time duration of Application/MPI phases for the single node benchmark of QE-CP-EU. The lighter zones identify higher point density.}
    \label{fig:freq_time}
\end{figure*}

\begin{figure*}[t]
    \centering
    \includegraphics[width=1.00 \linewidth]{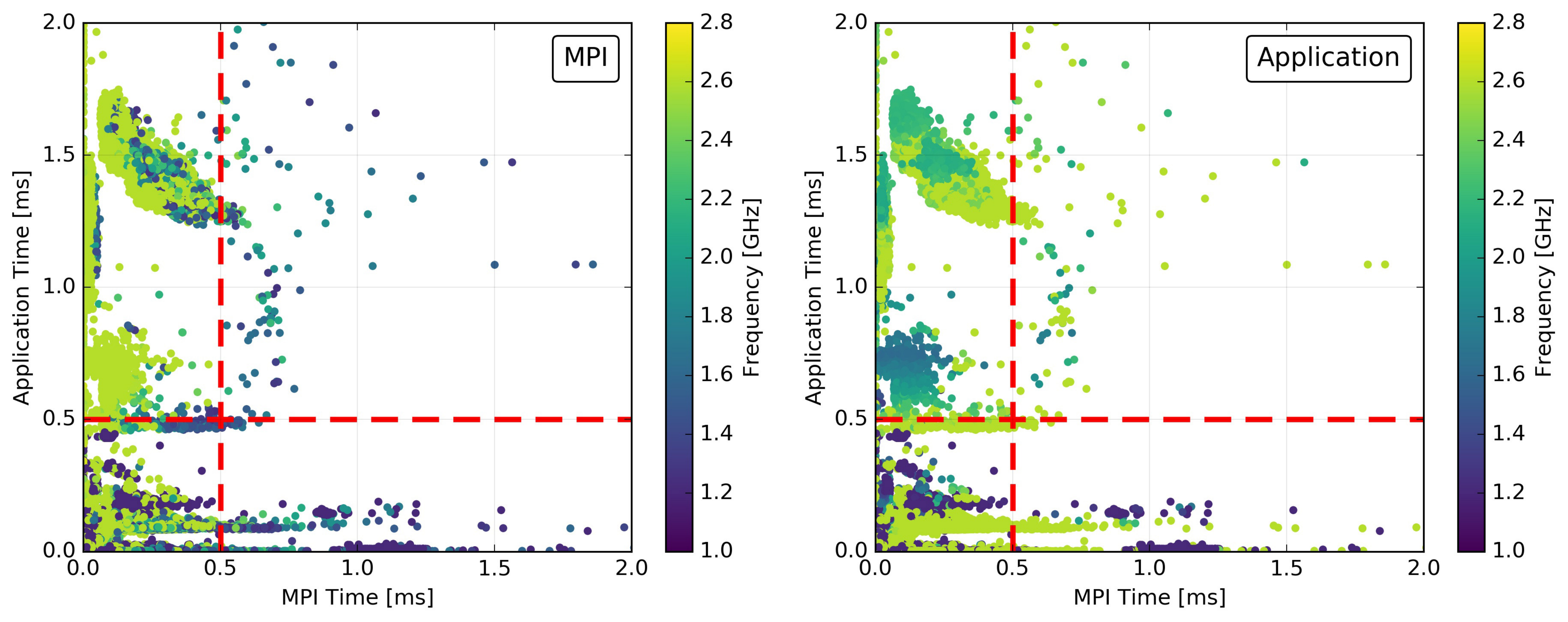}
    \caption{Time and average frequency of Application/MPI phases for the single node benchmark for QE-CP-EU.}
    \label{fig:time_time_freq}
\end{figure*}

\subsection{DVFS Overheads and Time Region Analysis}
%As we saw in Section \ref{sec:background}, it is not sufficient to simply instrument all the MPI primitives to scale down and up the frequency in each MPI primitive to reduce the energy consumption. We need to have a knowledge of the workload distribution and to the duration of the MPI communications to minimize the overheads and to achieve a good energy and power saving for the application.
%A HPC application can be seen as a composition of processes running on different nodes, where each process is composed by a sequence of application and MPI phases, one after the other, as depicted in figure \ref{fig:callback}.
To find the reason of the higher overhead when frequency reduction is applied in all the MPI phases as highlighted in the previous Section, we report two scatter plots in which we show on the x-axis of the left plot the time duration of each MPI phase and on the right plot the time duration of each application phase. For both plots, we report on the y-axes the measured average frequency in that phase. % for the single-node benchmark.
%We used the workload configuration where all MPI processes are involved in the diagonalization for the DVFS case.
%The x-axis represents the time duration of the phase, while the y-axis represents the average frequency in that phase.
This test is conducted by instrumenting each MPI call through COUNTDOWN with a \textit{prologue} routine to set the lowest frequency (1.2GHz) and with an \textit{epilogue} routine to set the highest frequency (Turbo).

In theory, we would have expected that all MPI phases had executed at the minimum frequency and application phases had always run at maximum frequency. 
It is a matter of fact that MPI phases running at high frequencies may cause energy waste, while application phases running at low frequencies cause a performance penalty to the application.
Our results show that for phases with a time duration between 0us and 500us, the average frequency vary in the interval between the high and low CPU's frequency values, while above it, it tends to the desired frequency for that phase.
This can be explained by the response time of \textit{HW power controller} in serving P-state transition of our Intel Haswell \cite{hackenberg2015energy}, we discover the same behaviour on Intel Broadwell architecture.
The \textit{HW power controller} periodically reads the DVFS register to check if the OS has specified a new frequency, this interval has been reported to be 500us in previous study \cite{hackenberg2015energy} and matches our empirical threshold.

This means that every new setting for the core's frequency faster than 500us could be applied or completely ignored, depending on when the register was sampled the previous time.
This can cause all sort of average frequencies. Clearly application phases which execute at a lower frequency than the maximum one may lead to a slowdown in the application, while MPI phases which execute at a higher frequency than the minimum one may lead to energy saving loss. It is nevertheless interesting to notice that phases with a duration from 0s to 500us are more likely to have the highest frequency for the MPI phases and the lowest frequency for the application phases. Which is the opposite of what expected. We will explain it with the next analysis. 
%The figure also shows that for the benchmark where all MPI processes are involved in the diagonalization, most of the MPI calls have a time duration less than 500us, which means that this benchmark is not a good candidate to apply our energy efficient strategy.

Thus, it is not possible to have effective control on the frequency selection for phases shorter than 500us, while for longer phases we have an asymptotic trend toward the requested frequency. We hypothesize that in phases shorter than 500us the average frequency depends more on the previous phase frequency than the requested one.  
%All MPI calls with a time duration less than 500us should be not considered for frequency scaling due to the impossibility to control.

%\begin{figure*}[t]
%    \centering
%    \includegraphics[width=1.00 \linewidth]{./images/cs_ps_ts_500us_n1_n16_v2}
%    \caption{Overhead, energy/power saving, average load and frequency when a single MPI process is involve in the diagonalization using COUNTDOWN. In the plot are depicted the C-state ($CS$), P-state ($PS$) and T-state ($TS$) mode. The overhead and the energy/power saving are compared with the busy-waiting mode (default mode) of MPI library.}
%    \label{fig:cs_ps_ts_500us_n1}
%\end{figure*}

\begin{figure*}
    \centering
    \captionsetup[subfigure]{position=top}
    \subfloat[All MPI processes diagonalization QE-CP-EU]{\includegraphics[scale=0.358]{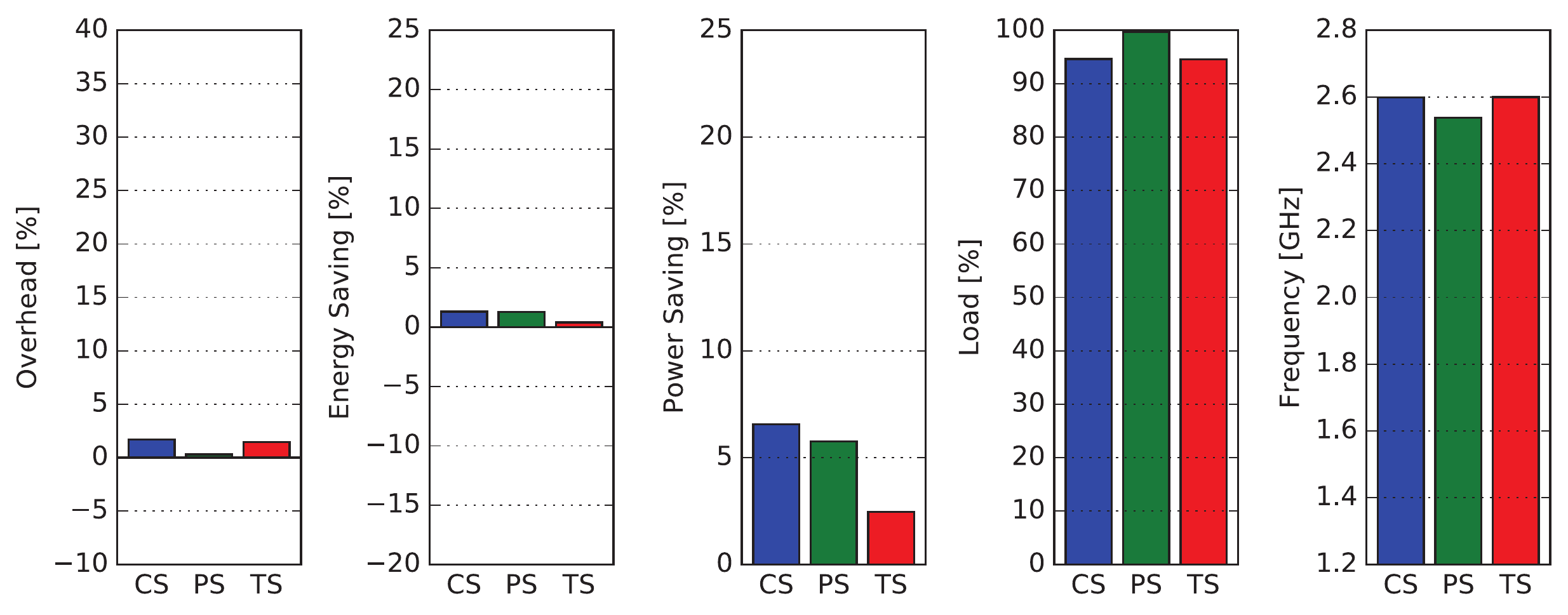} }
    \subfloat[Single MPI process diagonalization QE-CP-NEU]{\includegraphics[scale=0.358]{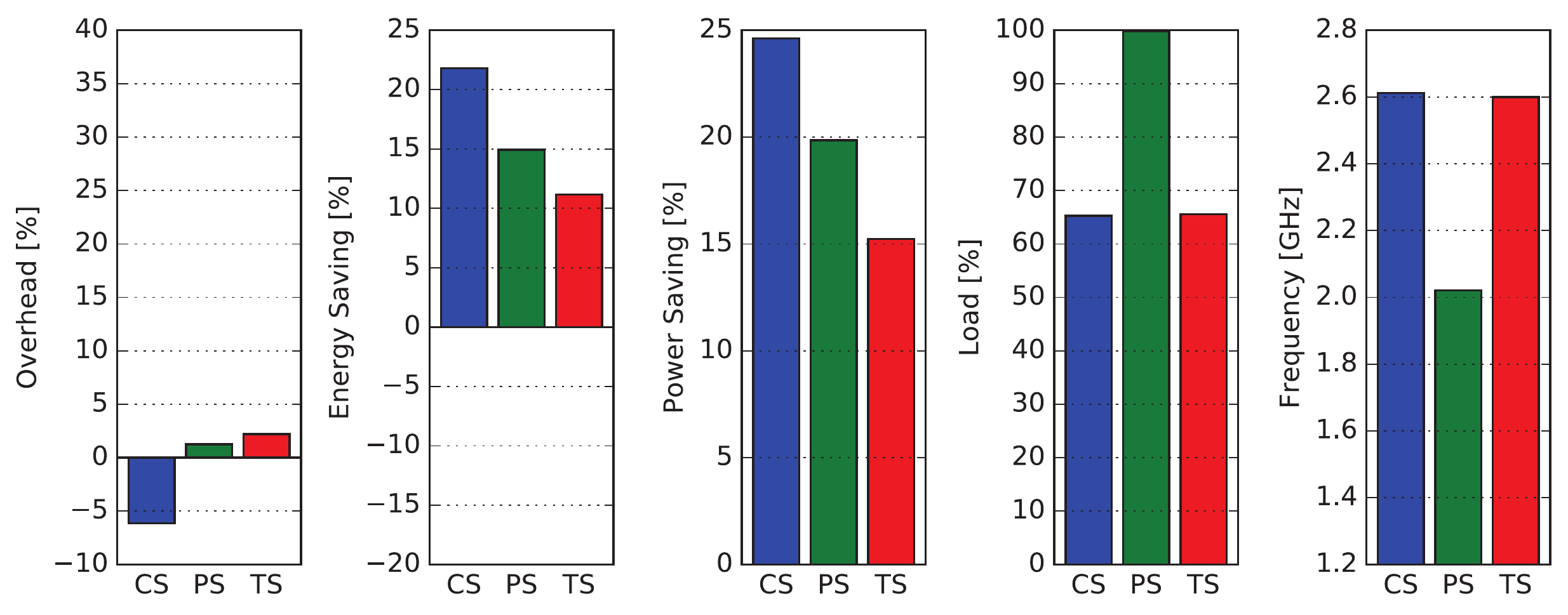} }
    \caption{Overhead, energy/power saving, average load and frequency using COUNTDOWN for QE-CP-EU (a) and QE-CP-NEU (b). Legend: C-state ($CS$), P-state ($PS$) and T-state ($TS$) mode. Baseline is busy-waiting mode of MPI library.}
    \label{fig:cs_ps_ts_500us}
\end{figure*}

Following this intuition, in Figure \ref{fig:time_time_freq}, we correlate the time duration of each application phase with the time duration of the following MPI phase and its average frequency.
We report in the y-axis the time duration of the application phase, in the x-axis the time duration of the subsequent MPI phase, and with the color code, we report the average frequency. In the left plot, we report the average frequency of the MPI phase, while in the right plot we report the average frequency for the application phase.
For both plots, we can identify four regions/quadrants:

%\begin{figure*}
%  \centering
%  \begin{subfigure}[b]{0.49\textwidth}
%    \centering
%    \textbf{(a) All MPI processes diagonalization QE-CP-EU}\par\medskip
%    \includegraphics[width=1.00\linewidth]{./images/cs_ps_ts_500us_n16}
    %\label{fig:cs_ps_ts_n16}
    %\caption{}
%  \end{subfigure}
%  ~
%  \begin{subfigure}[b]{0.49\textwidth}
%    \centering
%    \textbf{(b) Single MPI process diagonalization QE-CP-NEU}\par\medskip
%    \includegraphics[width=1.00\linewidth]{./images/cs_ps_ts_500us_n1}
    %\label{fig:cs_ps_ts_n1}
    %\caption{}
%  \end{subfigure}
%  \caption{Overhead, energy/power saving, average load and frequency using COUNTDOWN for QE-CP-EU (a) and QE-CP-NEU (b). Legend: C-state ($CS$), P-state ($PS$) and T-state ($TS$) mode. Baseline is busy-waiting mode of MPI library.}
%  \label{fig:cs_ps_ts_500us}
%\end{figure*}

%\begin{itemize}
(i) \textbf{Application \& MPI\textgreater 500us}: this region contains long application phases followed by long MPI phases. Points in this region show low frequency in MPI phases and high frequency in application phases. This is the ideal behavior, where applying frequency scaling policy reduces energy waste in MPI but with no impact on the performance of the application. Phases in this region are perfect candidates for fine-grain DVFS policies.

(ii) \textbf{Application\textgreater 500us \& MPI\textless 500us}: this region contains long application phases followed by short MPI phases. Points in this region show for both application and MPI phases high average frequency. This is explained by the short duration of the MPI phases, which does not give enough time to the \textit{HW power controller} to serve the request to scale down the frequency (\textit{prologue}) before this setting is overwritten by the request to operate at the highest frequency (\textit{epilogue}). For this reason, fine-grain DVFS control in this region does not have an impact on the energy saving as the frequency reduction in MPI phases is negligible, but it also does not deteriorate the performance as the application phases are executed at the maximum frequency. Phases in this region should not be considered for fine-grain DVFS policies, being preferable to leave frequencies unaltered at the highest level.
    
(iii) \textbf{Application\textless 500us \& MPI\textgreater 500us}: this region contains short application phases followed with long MPI phases. This is the opposite case of \textit{Application\textgreater 500us \& MPI\textless 500us} region. Points in this region show for both application and MPI phases low average frequency. This is explained by the short duration of the application phases, which does not give enough time to the \textit{HW power controller} to serve the request to raise up the frequency (requested at the exit of the previous MPI phase), before this setting gets overwritten by the request to operate at the lowest frequency (at the entrance of the following MPI phase). %This is explained by the fact that most of the time is spent in the MPI library and the HW power controller has not enough time to react in short application phases. %We can see that points that belong in this region show for both application and MPI phases high average frequency. 
Applying fine-grain DVFS policies in this region can save power, but detriments the overall performance, as application phases are executed at low frequencies. Phases in this region should not be considered for fine-grain DVFS policies due to the high overheads in the application execution time.
    
(iv) \textbf{Application \& MPI\textless 500us}: This region shows the opposite behavior of \textit{Application \& MPI\textgreater 500us} region. Both application and MPI phases execute randomly at high and low average frequencies due to the inability of the \textit{HW power controller} to capture and service the requested frequency changes. The average frequency at which MPI and application phases execute are strictly related to type of the previous long phase: if it was an application phase the following short phases will execute at high frequency in average; On the contrary, if it was an MPI phase the following short phases would execute at low frequency in average. Applying fine-grain DVFS policies in this region leads to unexpected behaviors which can detriment application performance. Fine-grain power managers should never consider all phases shorter than 500us.
%\end{itemize}

\begin{figure*}[t]
    \centering
    \includegraphics[width=\textwidth]{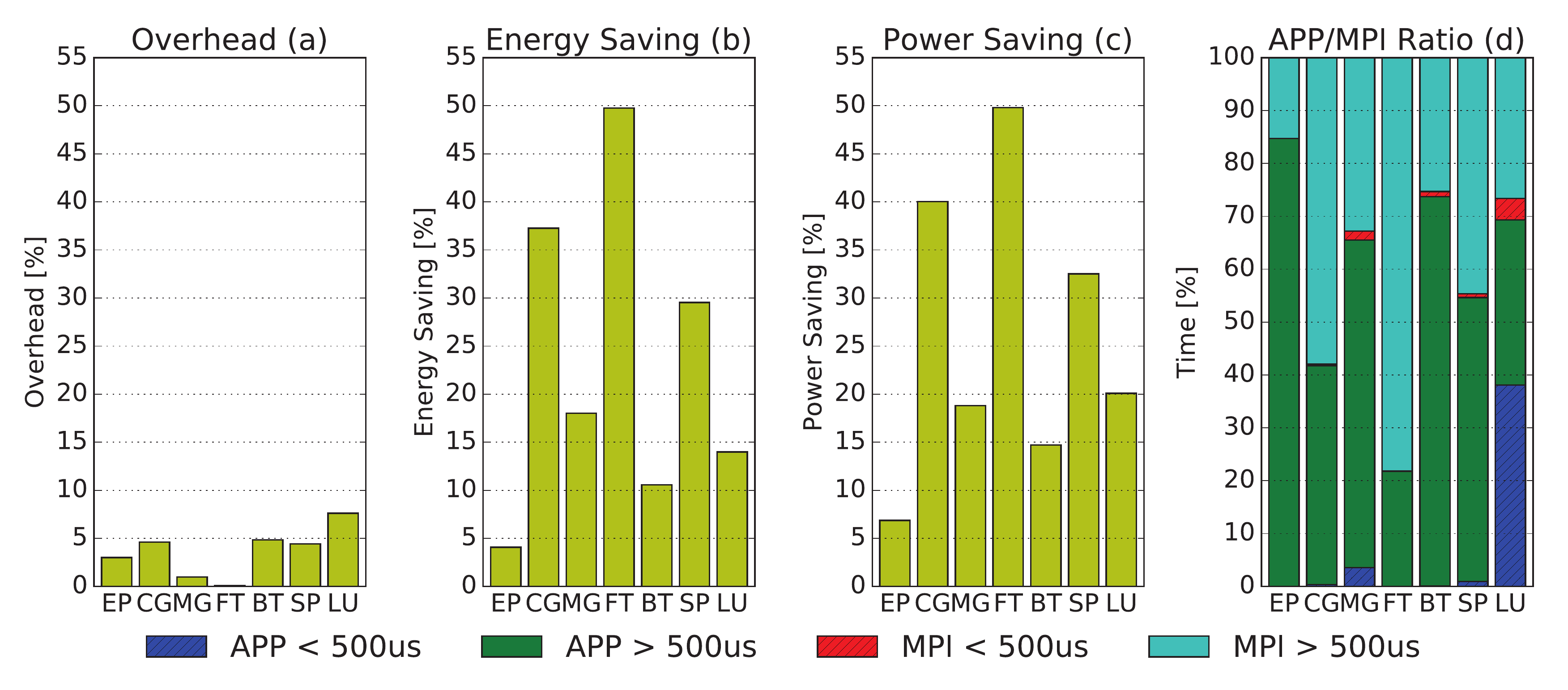}
    \caption{Results on NAS Benchmarks with COUNTDOWN. Baseline is busy-waiting mode (default mode) MPI library.}
    \label{fig:nas}
\end{figure*}

\subsection{Single-node Evaluation}
We repeated the experiments of Section \ref{sec:background} using COUNTDOWN.
We configure COUNTDOWN to scale down the P- and the T-states 500us after the prologues of MPI primitives.
%% DANIELE perchè T-state, non avevamo detto di non considerarli. -> secondo me è bene far vedere questi risultati perchè in questo modo valutiamo tutti i meccanismi utilizzabili per interagire con il power manager del sistema, inoltre citiamo l'articolo per utilizza i T-state quindi far vedere che sulle nuove generazioni di processori Intel il loro meccanismo non funziona altrimenti reviewer potrebbe dirci di utilizzare i T-state che potrebbero andare meglio ma noi facciamo vedere che non è vero

To reproduce the same timeout strategy leveraging the C-states, we configure \textit{MPI SPIN WAIT} as described in \ref{sec:event_module} with 10K as MPI spin counter parameter. 

%on the idleness of the system instead of the DVFS knobs, we use the capability of Intel MPI to configure the number of times to spin in an MPI primitive before to release the execution to the idle task of the CPU, as we described in \ref{sec:event_module}.
The \textit{HW power controller} of Intel CPUs, has a different transition latency for sleep states w.r.t. DVFS scaling, as described in \cite{hackenberg2015energy}.
For this reason, we empirically determine the best spin counter setting to maximize energy efficiency and to minimize the overhead for the target application. %\footnote{
%Thus, we run QE varying this MPI spin counter parameter with logarithmic steps in an interval between 1 to 1 Billion.
%We found that 10K shows the best results in term of energy saving and overhead. We omit the results for all the steps for lack of space and we use this setting in the next comparison.}.

Figure \ref{fig:cs_ps_ts_500us} report the experimental results using \textit{COUNTDOWN THROTTLING}, \textit{COUNTDOWN DVFS} and \textit{MPI SPIN WAIT}. %the Intel MPI library with spincounter configuration.
We can see that in all cases the overhead, the energy saving, and the power saving are significantly improved w.r.t. the baseline (only MPI library).

Figure \ref{fig:cs_ps_ts_500us}.a shows the experimental results for QE-CP-EU.
%\textbf{DANIELE} \tetit{Non capisco un tubo di quello che dici dopo. Che vuol dire "is down from"}
For the C-state mode the overhead decrease from 25.85\% to 1.70\% by using \textit{MPI SPIN WAIT}.
Instead, for the P-state using \textit{COUNTDOWN DVFS} the overhead decreases from 5.96\% to a negligible overhead, and for the T-state using \textit{COUNTDOWN THROTTLING} the overhead decreases from 5.96\% to 0.29\%.
All evaluations report a non-negative energy saving, as it was for the MPI library without timeout strategy, but with better results.
Energy saving shows 21.80\%, 14.94\%, and 11.16\% improvements and power saving report 6.55\%, 5.77\%, and 2.47\% respectively for C-state, P-state, and T-state.
These experimental results confirm our exploration of the time duration of MPI phases reported in figure \ref{fig:freq_time}. Most of the MPI calls of this benchmark have been skipped due to their short duration to avoid overheads.

Figure \ref{fig:cs_ps_ts_500us}.b show similar improvements for QE-CP-NEU.
In this configuration, for C-State mode the speed-up increases from 1.08\% to 6.14\% using \textit{MPI SPIN WAIT}.
Instead of using COUNTDOWN, the overhead of P-state decreases from 3.88\% to 1.25\%, and for the T-state from 15.82\% to 2.19\%.
As a result, the energy saving is 21.80\%, 14.94\%, and 11.16\% while power saving corresponds to 24.61\%, 19.84\%, and 15.23\% respectively for C-state, P-state, and T-state.

\subsection{HPC Evaluation}
After we have evaluated our methodology in a single compute node, we extend our exploration in a real HPC system.
%% MOVED FROM BACKGROUND
We use a Tier-1 HPC system based on an IBM NeXtScale cluster which is currently classified in the Top500 supercomputer list \cite{TOP500}.
The compute nodes of the HPC system, are equipped with 2 Intel Broadwell E5-2697 v4 CPUs, with 18 cores at 2.3 GHz nominal clock speed and 145W TDP and interconnected with an Intel QDR (40Gb/s) Infiniband high-performance network.

%%% APPLICAZIONI + NAS + QE

To benchmark the parallel performances in our target HPC system we focused on two set of applications. The first one is the NAS parallel benchmark suite \cite{nas} with the dataset E. We executed the NAS parallel benchmarks on 29 compute nodes with a total core count of 1024 cores. We use 1024 cores due the execution time of the application run using dataset E is on average ten minutes for each benchmark.
The second one is the QuantumESPRESSO PWscf software configured for a complex large-scale simulation. For this purpose, we performed ten iterative steps of the self-consistent loop algorithm that optimizes the electronic density starting from the superposition of atomic charge densities. To obtain a reasonable scaling up to the largest set of nodes, we chose an ad-hoc dataset.
%The selected system comes from an actual scientific report \cite{avvisati2017fepc} and reproduces a layered structure of Iridium, Cobalt, and Graphene plus a molecular compound (iron-phthalocyanine) deposited on top. The whole simulation box includes 662 atoms which are described with 3662 KS states. The total number of plane waves is more than a million and, during the execution, main memory occupation may peak at 2 to 6 terabytes depending on the selected parallelization parameters.

%\begin{figure}[t]
%    \centering
%    \includegraphics[width=1.00 \linewidth]{./images/cs_ps_ts_n1}
%    \caption{Overhead, energy/power saving, average load and frequency when only a MPI process participate to the diagonalization. Legend: C-state ($CS$), P-state ($PS$) and T-state ($TS$) mode. Baseline is busy-waiting mode (default mode) of MPI library.}
%    \label{fig:cs_ps_ts_n1}
%\end{figure}

During each iteration, the CPU time is mostly spent in linear algebra (matrix-matrix multiplication and matrix diagonalization) and FFT. Both these operations are distributed on multiple processors and operate on distributed data. As a consequence, FFT requires many AllToAll MPI communications while parallel diagonalization, performed with the \textsc{PDSYEVD} subroutine of \textsc{ScaLAPACK} and requires mostly MPI broadcasting messages. 
We run QE on 96 compute nodes, using 3456 cores and 12 TB of DRAM due our target HPC machine allows application runs with at maximum 100 nodes.
We use an input dataset capable of scaling on such number of cores, and we configure QE using a set of parameters optimized to avoid network bottlenecks, which would limit the scalability. We name this configuration QuantumESPRESSO Expert User (QE-PWscf-EU), to differentiate it from the same problem but solved without optimizing the internal parameter as it was run by a user without domain-specific knowledge which we call QuantumESPRESSO Not Expert User (QE-PWscf-NEU). 

%\begin{figure*}
%  \centering
%  \begin{subfigure}[b]{0.38\textwidth}
%    \centering
%    \textbf{(a) QE-PWscf-EU}\par\medskip
%    \includegraphics[width=1.00\linewidth]{./images/multi_node}
%    %\label{fig:cs_ps_ts_n16}
%    %\caption{}
%  \end{subfigure}
%  ~
%  \begin{subfigure}[b]{0.21\textwidth}
%    \centering
%    \includegraphics[width=1.00\linewidth]{./images/oep_multi_node}
%    %\label{fig:cs_ps_ts_n16}
%    %\caption{}
%  \end{subfigure}
%  ~
%  \begin{subfigure}[b]{0.38\textwidth}
%    \centering
%    \textbf{(b) QE-PWscf-NEU}\par\medskip
%    \includegraphics[width=1.00\linewidth]{./images/multi_node_unbalanced}
%    %\label{fig:cs_ps_ts_n1}
%    %\caption{}
%  \end{subfigure}
%  \caption{(a,b) Sum of the time spent in phases longer and shorter than 500us for QE-PWscf-EU and QE-PWscf-NEU.}
%  \label{fig:multi_node}
%\end{figure*}

\begin{figure*}
    \centering
    \captionsetup[subfigure]{position=top}
    \subfloat{{\includegraphics[width=0.39\linewidth]{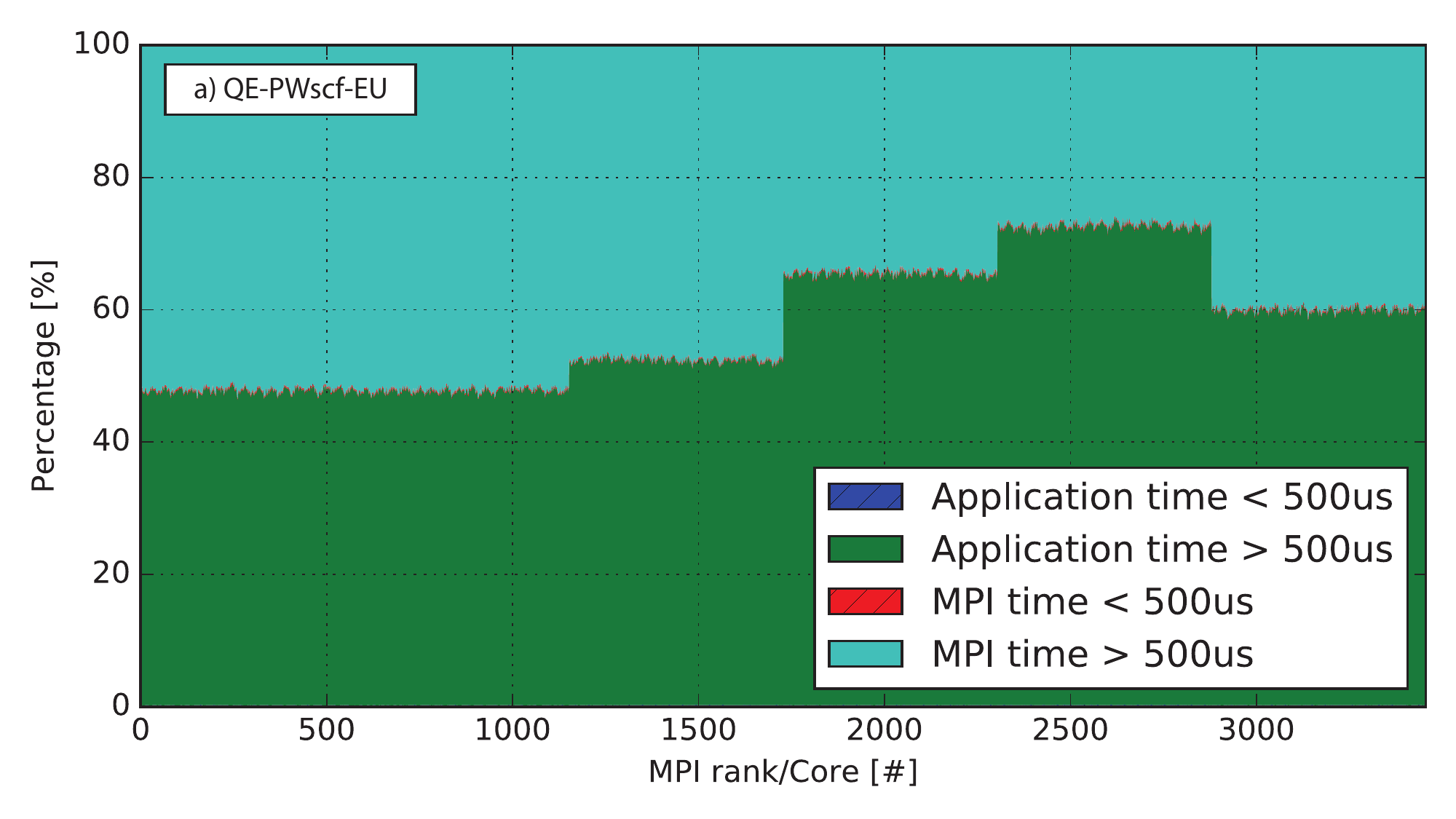} }}
    \subfloat{{\includegraphics[width=0.22\linewidth]{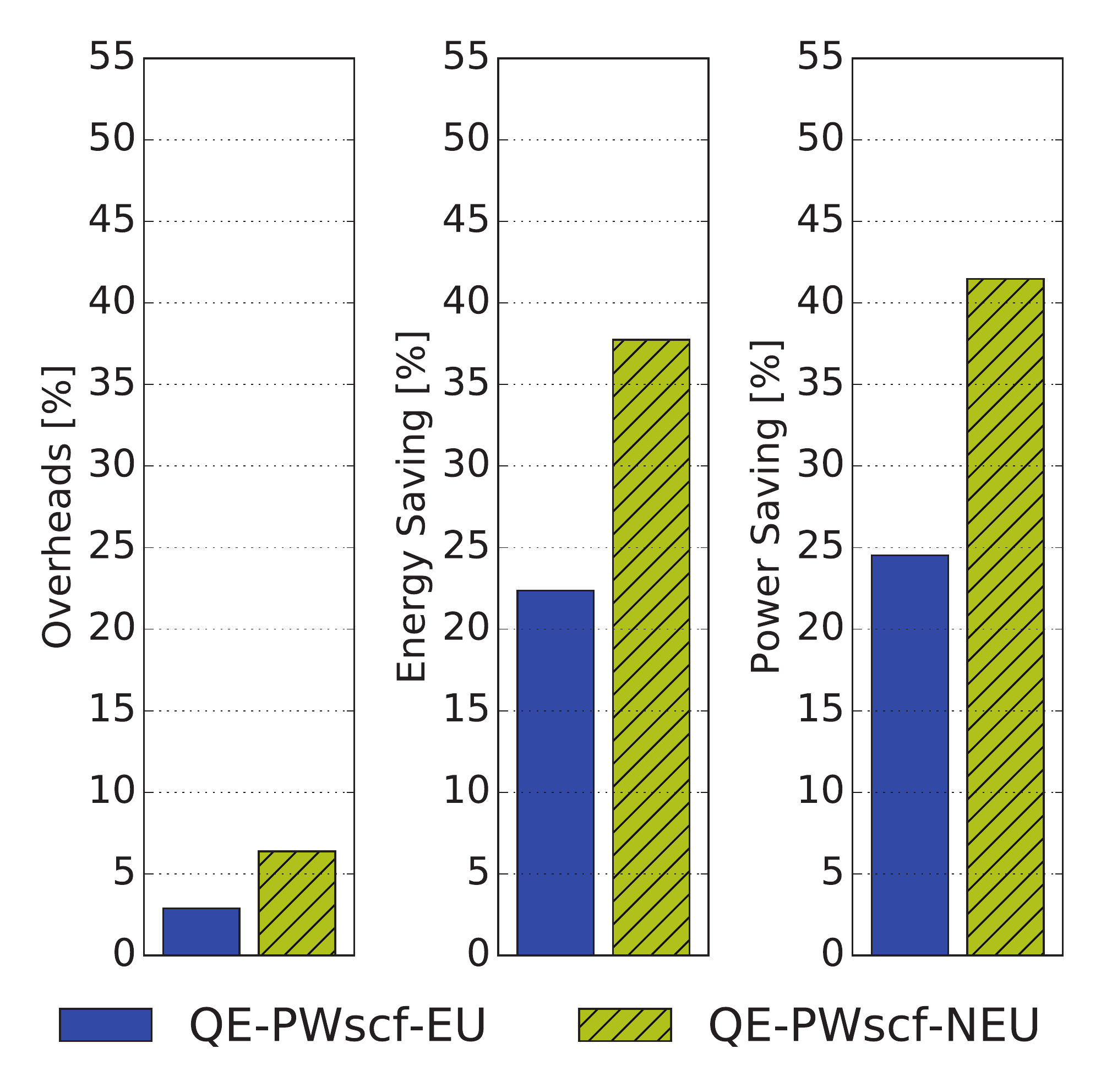} }}
    \subfloat{{\includegraphics[width=0.39\linewidth]{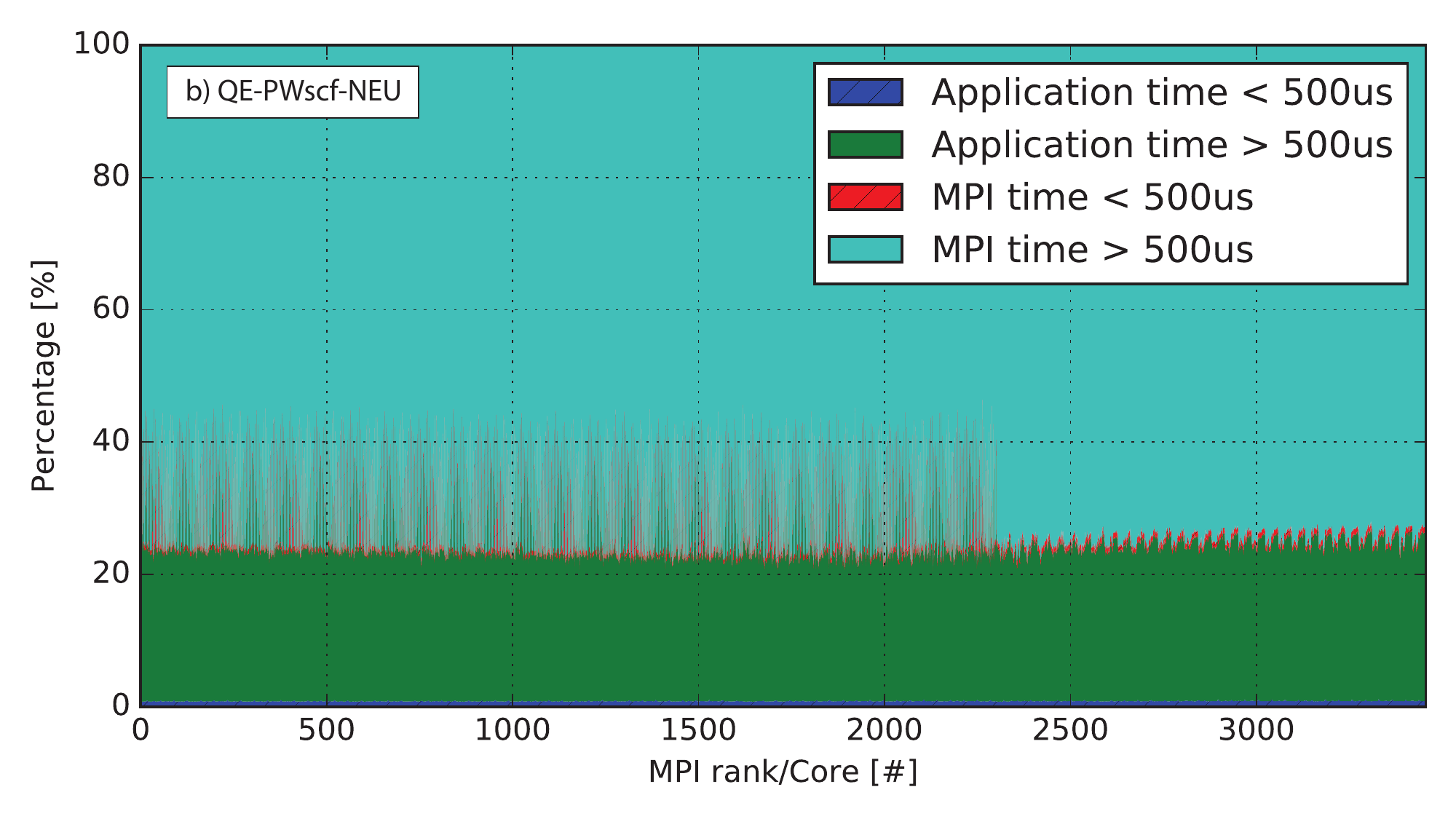} }}
    \caption{(a,b) Sum of the time spent in phases longer and shorter than 500us for QE-PWscf-EU and QE-PWscf-NEU.}
    \label{fig:multi_node}
\end{figure*}

In these tests, we exclude the T-state mode, because, in the single-node evaluation, it always reported the worst results that the P-state mode.
We also excluded the C-state mode as when we started the configuration of the Intel MPI library for HPC experiments using idle mode. We discover that this feature is not supported in a distributed environment.
The Intel MPI library overrides the request of idle mode with the busy-wait mode when the application runs on multiple nodes.
For this reason, we only use the P-state mode (\textit{COUNTDOWN DVFS}) in the HPC evaluation.

We run the benchmark with and without COUNTDOWN on the same nodes, and we compared the results.

Figure \ref{fig:nas} shows the results for the NAS parallel benchmark suite when executed on 1024 cores, while figure \ref{fig:multi_node} shows the results for the QE-PWscf-* application when executed on 3456 cores. The different plots for Figure \ref{fig:nas} reports the time-to-solution overhead, the energy and power saving as well as the MPI and application time phases distribution (in the percentage of the total time the accumulated time spent in phases longer and shorter than 500us) for the different large-scale benchmarks and application run. All the values are normalized against the default MPI busy waiting policy. From Figure \ref{fig:nas}.c, we can see that COUNTDOWN is capable of significantly cutting the energy consumption of the NAS benchmarks from 6\% to 50\%. From Figure \ref{fig:nas}.c we can see that this savings follows the percentage of time the benchmark passes in MPI phases longer than 500us. From the overhead plot (Figure \ref{fig:nas}.a) we can see that all these energy savings happen with a very small time-to-solution overhead, on average below 5\%. These results are very promising as they are virtually portable to any application, without the need to touch the application binary. When looking at the QuantumEspresso (QE-PWscf-*) case reported in figure \ref{fig:multi_node}, we see that COUNTDOWN attains similar results of NAS also with real production run optimized for scalability COUNTDOWN saves 22.36\% of energy with an overhead of 2.88\% in the QE-PWscf-EU case.

Figure \ref{fig:multi_node}.a shows the total time spent in the application and in MPI phases which are shorter and longer than 500us for the QE-PWscf-EU case.
On the x-axis, the figure reports the Id of the MPI rank, while in the y-axis reports in the percentage of the total time spent in phases longer and shorter than 500us.
We can immediately see that in this real and optimized run, the application spends a negligible time in phases shorter than 500us.
In addition, the time spent in the MPI library and the application is not homogeneous among the MPI processes.
This is an effect of the workload parameters chosen to optimize the communications, which distribute the workload in subsets of MPI processes to minimize broadcast and All-to-All communications.
Using this configuration, our experimental results report 2.88\% of overhead with an energy saving of 22.36\% and a power saving of 24.53\% thanks to COUNTDOWN.

%\begin{figure}[t]
%    \centering
%    \includegraphics[width=1.00 \linewidth]{./images/multi_node}
%    \caption{Sum of the time spent in phases longer and shorter than 500us for the QE-PWscf-EU.}
%    \label{fig:multi_node}
%\end{figure}

%\begin{figure}[t]
%    \centering
%    \includegraphics[width=1.00 \linewidth]{./images/multi_node_unbalanced}
%    \caption{Sum of the time spent in phases longer and shorter than 500us for the benchmark without the optimization for the communications.}
%    \label{fig:multi_node_unbalanced}
%\end{figure}

Figure \ref{fig:multi_node}.c shows that for the case QE-PWscf-NEU where the parameters are not optimized, all MPI processes have the same workload composition as they are part of the same workgroup and due the large overhead in the broadcast and All-to-All communications. Most of the processes spend almost 80\% of the time in the MPI library.
Even if it is suboptimal, this happens to HPC users running the application without being domain experts or before tuning the execution parameters. This is a rather typical scenario in scientific computing as only runs that are repeated multiple times are carefully optimized by domain experts.

In this situation, COUNTDOWN increases its benefits, reaching up to 37.74\% of energy saving and a power saving of 41.47\%.
In this condition, we also notice that COUNTDOWN induces a small but relevant overhead of 6.38\%.
We suspect that some MPI primitives suffer more than others from the frequency scaling. We will analyze in depth this problem in our future works aiming to guarantee that the COUNTDOWN overhead always remains negligible. However, we remark that an overhead well below 10\% is more than acceptable in many HPC facilities, especially when considering the massive energy savings.  

In summary, we can conclude that results achieved by COUNTDOWN in at production scale and application are very promising and if systematically adopted would dramatically reduce the TCO of today supercomputers.

%% file: conclusion.tex
\section{Conclusion}
\label{sec:conclusion}

In this paper, we presented COUNTDOWN, a methodology and a tool for profiling HPC scientific applications and for adding DVFS capabilities into standard MPI libraries.
COUNTDOWN implements a timeout strategy to avoid application slowdown and exploiting MPI communication slacks to reduce energy consumption drastically.
COUNTDOWN has been demonstrated on real HPC systems and workloads and does not require any modification to application source code nor the compilation toolchain. The COUNTDOWN approach can leverage several low power state technologies --- P/T/C states.

We compared COUNTDOWN with state-of-the-art power management approaches for MPI libraries, which can dynamically control idle and DVFS levels for MPI-based application.
Our experimental results show that using our tuned timeout strategy to take decisions on power control can drastically reduce overheads, maximizing the energy efficiency in small and large MPI communications.
Our run-time library can lead up to 14.94\% energy saving, and 19.84\% of power saving with a less than 1.5\% performance penalty on a single compute node.
However, the benefits of COUNTDOWN increase with the scale of the application. In a 1K cores NAS run, COUNTDOWN always saves energy, with a saving which depends on the application and ranges from 6\% to  50\% at a negligible overhead (below 6\%). In a full-scale production run of QE on more than 3.4K cores, COUNTDOWN saves 22.36\% of energy with only 2.88\%  performance overhead. Energy reduction reaches 37.74\% when the application is executed with a default conservative parallelization setting. 

COUNTDOWN is an effective, non-intrusive and low overhead approach to cut today's supercomputing center energy-consumption transparently to the user. In future work, we plan to integrate it within standard power management infrastructure, such as GEOPM \cite{GEOPM}, and to complement it with predictive and application-driven power management techniques.